\documentclass{aa}
\usepackage{epsf}
\newcommand{\msol}{M_\odot}

\newcommand{\halpha}{H$\alpha$}
\newcommand{\hbeta}{H$\beta$}
\newcommand{\hgamma}{H$\gamma$}
\newcommand{\hdelta}{H$\delta$}
\newcommand{\hepsilon}{H$\epsilon$}

\newcommand{\teff}{T_{\mathrm{eff}}}         
\newcommand{\nh}{n_{\mathrm H}}             
\newcommand{\unt}[1]{_{\mathrm{#1}}} 
\def\fm{\hbox{$.\!\!^{\mathrm m}$}}
\def\mag{\hbox{$\hbox{\thinspace \fm}$}}

\newcommand{\nhe}{n\unt{He}}

\newcommand{\litanf}{\begin{list}{}{\leftmargin=1.5cm \rightmargin=0cm 
\itemindent=-1.5cm \parsep=0cm \itemsep=0cm }}
\newcommand{\litend}{\end{list}}

\newcommand{\aj}{AJ}
\newcommand{\apj}{ApJ}
\newcommand{\apjs}{ApJS}

\newcommand{\aua}{A\&A}
\newcommand{\auas}{A\&AS}
\newcommand{\mnras}{MNRAS}

\newcommand{\pasp}{PASP}

\begin{document}

\thesaurus{06(08.01.1; 08.06.3; 08.23.1; 09.16.1)}

\title{Spectroscopic investigation of old planetaries}
\subtitle{IV.~Model atmosphere analysis}
\author{R.~Napiwotzki
\thanks{Visiting astronomer, German-Spanish Astronomical Center, Calar Alto,
Spain, operated by the Max-Planck-Institut f\"ur Astronomie,
Heidelberg, jointly with the Spanish National Commission for Astronomy}
}
\institute{
Dr.~Remeis-Sternwarte, Sternwartstr.~7, D-96049~Bamberg, Germany
}
\offprints{Ralf Napiwotzki (napiwotzki@sternwarte.uni-erlangen.de)}
\date{Received date; accepted date}

\maketitle
\markboth{R.~Napiwotzki: Spectroscopic investigation
	of old planetaries  IV.}
	{R.~Napiwotzki: Spectroscopic investigation
	of old planetaries  IV.}

\begin{abstract}
The results of a NLTE model atmosphere analysis of 27 hydrogen-rich central
stars of old planetary nebulae (PN) are reported. These stars were selected
from a previous paper in this series, where we gave classifications for a total
of 38 central stars. Most of the analyzed central stars fill a previously
reported gap in the hydrogen-rich evolutionary sequence. Our observations imply
the existence of two separated spectral evolutionary sequences for
hydrogen-rich and -poor central stars/white dwarfs. This is in line with
theoretical evolutionary calculations, which predict that most post-AGB stars
reach the white dwarf domain with a thick hydrogen envelope of $\approx
10^{-4}\msol$. 

We determine stellar masses from the comparison with evolutionary tracks and
derive a mass distribution for the hydrogen-rich central stars of old PNe. The
peak mass and the general shape of the distribution is in agreement with recent
determinations of the white dwarf mass distribution. 

The properties of most analyzed stars are well explained by standard post-AGB
evolution. However, for eight stars of the sample other scenarios have to be
invoked. The properties of three of them are probably best explained by born
again post-AGB evolution. Two of these are hybrid CSPN (hydrogen-rich PG\,1159
stars), but surprisingly the third star doesn't show any signs of chemical
enrichment in its atmosphere. The parameters of five stars are not in
accordance with post-AGB evolution. We discuss alternative scenarios such as
the stripping of the hydrogen-rich envelope by a companion during the first red
giant phase or the formation of a common envelope with a possible merging of
both components. Two stars (\object{HDW\,4} and \object{HaWe\,5}) remain
mysterious after all. They resemble ordinary hot DA white dwarfs, but due to
very large evolutionary ages the presence of a PN cannot be explained. We
speculate that the nebulae may be shells produced by ancient nova outbursts. 

A wide spread of helium abundances is observed in the photospheres of central
stars of old PNe. It is shown that a good correlation between helium abundances
and luminosity is present. It is inferred that when the stars' luminosities
fall below $L\approx 300 L_\odot$ depletion starts and the helium abundance
steadily decreases with decreasing luminosity. The existence of this
correlation is in qualitative agreement with recent theoretical calculations of
gravitational settling in the presence of a stellar wind. 


\end{abstract}

\section{Introduction}
Central stars of planetary nebulae (PN) are the immediate precursors of white
dwarfs. Most stars enter the white dwarf cooling sequence through this
evolutionary channel (Drilling \& Sch\"onberner \cite{DS85}). After the maximum
effective temperature is reached (100,000\,K and more) the nuclear burning
ceases, the surrounding nebula disperses, and the central star stage ends. Thus
the nuclei of old PNe mark the transition to the white dwarfs. During this
stage the onset of gravitational settling, which causes the chemical purity of
many white dwarf atmospheres, can be observed. 

Central stars of old PNe play a crucial role in our understanding of the
formation of the two distinct sequences of central stars of planetary nebulae
(CSPN) and white dwarfs: the hydrogen-rich and the hydrogen-deficient (helium-
and carbon-rich) one. For the hydrogen-deficient CSPN there exists a continuous
sequence from the luminous central stars (Wolf-Rayet stars of spectral type
[WC]) to the low luminosity DO white dwarfs via the PG\,1159 stars, hot
pre-white dwarfs with typical temperatures of 100000\,K. M\'endez (\cite{M91})
presented a compilation of spectral types for (mostly luminous) CSPN and found
a fraction of about 1/3 hydrogen-deficient objects. Liebert (\cite{L86})
reported a ratio of H-poor to H-rich stars of 1:(7$\pm$3) for the hot white
dwarfs (spectral types DO and DA resp.) in the Palomar-Green survey (Green et
al.\ \cite{GSL86}), but noted a lack of very hot DA white dwarfs (the
counterparts of the PG\,1159 stars). Holberg (\cite{H87}) selected a sample of
the apparently hottest DA/DAO white dwarfs from the PG survey and concluded
that none of them appear to have a temperature in excess of 80000\,K, with
$T_{\mathrm{eff}}$ typically in the range 60000\,K to 70000\,K. However,
detailed analyses of these stars were not presented. 

Fontaine \& Wesemael (\cite{FW87}) discussed this seeming lack of very hot
H-rich white dwarfs and proposed that all CSPN finally lose their hydrogen-rich
surface layer and evolve through one channel: the PG\,1159/DO stars. After the
stars have become white dwarfs gravitational settling causes small traces of
hydrogen, hidden in the helium layer, to float up and form a very thin H-layer.
This scenario is in contradiction to theoretical calculations (e.g.\
Sch\"onberner \cite{S81}, \cite{S83}; Vassiliadis \& Wood \cite{VW94};
Bl\"ocker \cite{B95}), which predict a ``thick'' H-layer of
$\approx$$10^{-4}\msol$, but could nicely explain many properties of white 
dwarfs known at that time (EUV/X-ray observations, the DB gap, pulsational 
properties of ZZ\,Ceti stars). Due to improved observations and increased 
theoretical knowledge the interpretation has changed during the last years and 
it is now generally accepted that the vast majority of DA white dwarfs possess 
a thick H-layer (cf.\ the recent review of Fontaine \& Wesemael \cite{FW97}). 
However, the interpretation of many observational results is not completely 
unambiguous. Vennes et al.\ (\cite{VTG97}), e.g., analyzed a sample of EUV 
selected DA white dwarfs and compared their results with cooling tracks 
calculated with thick and with very thin layers. They found better agreement 
with the latter tracks and interpreted this result as evidence of a very thin 
hydrogen layer of most DA white dwarfs (but see discussion in Napiwotzki 
et al.\ \cite{NGS99}). 

The Fontaine \& Wesemael (\cite{FW87}) scenario makes one prediction concerning
the stars just entering the white dwarf cooling sequence: the vast majority of
them should be hydrogen-poor PG\,1159 stars or DO white dwarfs. Although a
small number of hydrogen-rich stars in this transition region don't contradict
the very thin layer scenario (see discussion in Fontaine \& Wesemael
\cite{FW97}), the hydrogen-poor ones should be the dominant species. The
transition between CSPN and white dwarfs is represented by the central stars of
old PNe. This prompted us to start a systematic survey of the nuclei of old
PNe. We selected apparently very old, faint and extended nebulae whose central
stars could be identified. Important sources were the article of Kwitter et
al.\ (\cite{KJL88}) and the compilation of nearby PNe by Ishida \& Weinberger
(\cite{IW87}). The central stars should be far evolved, just entering the white
dwarf cooling sequence. First results on interesting stars were reported in
Sch\"onberner \& Napiwotzki (\cite{SN90}; Paper I) and Napiwotzki \&
Sch\"onberner (\cite{NS91a}; Paper II). A spectral classification of the
complete sample of 38 CSPN is provided in Napiwotzki \& Sch\"onberner
(\cite{NS95}; Paper III). It turned out that 28 central stars have a
hydrogen-rich surface composition and that only seven were hydrogen-poor. The
ratio of hydrogen- rich and -poor objects in our sample of central stars just
entering the white dwarf sequence amounts to 4:1, in good agreement with the
previous investigations of Liebert (\cite{L86}) and M\'endez (\cite{M91})
mentioned above. Our result indicates that the ratio of H-rich and H-poor
objects remains approximately constant during the evolution from the central
star phase to the white dwarf cooling sequence, contrary to what is expected
from the Fontaine \& Wesemael (\cite{FW87}) scenario. Since this finding is of
fundamental meaning for our understanding of many aspects of white dwarf
evolution, we will present here the results of a model atmosphere analysis of
our central stars. This will allow us to determine their precise evolutionary
status. We will in this paper concentrate on the hydrogen-rich central stars,
which are the key for our understanding of the pre-white dwarf evolution.
Analyses of hydrogen-poor Wolf-Rayet CSPN were recently carried out by
Leuenhagen et al.\ (\cite{LHJ96}) and Koesterke \& Hamann (\cite{KH97}). A
review on the properties of PG\,1159 stars is given by Dreizler et al.\
(\cite{DWH95}). Prior to our survey only three analyses of high gravity CSPN
were published: \object{A\,7} (M\'endez et al.\ \cite{MKG81}),
\object{NGC\,7293} (M\'endez et al.\ \cite{MKH88a}), \object{EGB\,6} (Liebert
et al.\ \cite{LGB89}). Our much enlarged sample now allows a more meaningful
investigation of the stellar evolution in this region of the HR diagram. 

This paper is organized as follows. We start with a description of the
observations and their analysis in Sect.~2 and ~3, respectively. The results of
the model atmosphere analysis are presented and the evolutionary status of the
CSPN is discussed in Sect.~4. The evolution of the helium abundance of
hydrogen-rich central stars and white dwarfs is described and compared to
theoretical expectations in Sect.~5. The article finishes with conclusions in
Sect.~6. We will use the results presented in this article to discuss the
question of the PNe distance scale in Paper~V of this series. 

\section{Observations and data reduction}
A description of the sample selection and observational details was already
given in Paper~III. Therefore we will provide here only a cursory overview and
details on additional observations. 

Spectroscopic observations were performed with the 3.5\,m telescope of the
Calar Alto Observatory/Spain and the Cassegrain TWIN spectrograph. We performed
three runs with low resolution set-ups (October 1989, September 1990, and July
1991). Medium resolution follow-up observations were performed in July 1992,
July 1995, and October 1995. The resolution of the low resolution spectra
ranges from 5 to 8\,\AA. Details on the first four runs are provided in
Paper~III. 

The July 1995 and the October 1995 runs used identical set-ups. Grating T\,05
in the blue and T\,09 in the red yielded a dispersion of 36\,\AA/mm. Tek CCDs
\#11 and \#12 with $1024 \times 1024$ pixels of size 22\,$\mu$m were used. The
resulting spectral resolution, as measured from the He and Ar lines of the
comparison spectra, amounts to 1.5\ldots 1.8\,\AA\ (FWHM). While we always
exposed the region from 5600\,\AA\ to 6800\,\AA\ in the red, we needed two
grating angle settings to cover the spectral range from 3700\,\AA\ to
5400\,\AA\ in the blue chanel. However, we did not always get spectra for both
ranges. The data reduction follows standard procedures and is identical to that
described in Paper~III. Let us only remark that we tried to correct for the
nebula (and night sky) emission lines. For PNe with moderate line strengths and
smooth density distributions this worked well. In cases of PNe with high
surface brightness and/or strong spatial inhomogenities contamination by
nebular emission lines remained (either insufficient subtraction or
overcorrection). Examples are \object{NGC\,6720} and \object{HaWe\,5}. In these
cases the cores of the stellar Balmer lines (\halpha\ and \hbeta\ most
severely) and the \ion{He}{ii} 4686\,\AA\ line may be corrupted. 

We have added the CSPN \object{PHL\,932} and the hot DAO white dwarf
\object{HZ\,34} from Bergeron et al.\ (\cite{BWB94}) to our original sample
(Paper~III). On the other hand we didn't analyse the spectra of
\object{Sh\,2-176} and \object{IW\,2} from Paper~III, because their quality is
just good enough for a reliable classification but is too poor for a reasonable
analysis. In the case of \object{WeDe\,1} (\object{WDHS\,1}) we based our
analysis on the spectrum taken by Liebert et al.\ (\cite{LBT94}), because it's
signal-to-noise ratio is higher than that of our spectrum. 

\section{Model atmosphere calculations and fit procedure}

We calculated hydrogen and helium composed model atmospheres with the NLTE code
developed by Werner (\cite{W86}). Basic assumptions are those of static,
plane-parallel atmospheres in hydrostatic and radiative equilibrium. In
contrast to the atmospheres commonly used to analyze DA white dwarfs, we relax
the assumption of local thermal equilibrium (LTE) and solve the detailed
statistical equilibrium instead of the Saha-Boltzmann equations. As described
in Werner (\cite{W86}), the accelerated lambda iteration (ALI) method is used
to solve the set of non-linear equations. 

The analysis of white dwarfs is usually the domain of model atmospheres in LTE.
It is assumed that deviations from LTE are kept small by the high densities in
white dwarf atmospheres. Model atmospheres which drop the LTE assumption are
somewhat loosely called non-LTE (NLTE) atmospheres. A detailed check of the
validity of the LTE assumption was presented in Napiwotzki (\cite{N97}). It
could be shown that LTE is a very good approximation for the ``classical'' DA
white dwarfs with temperatures below  40000\,K. However, the situation is
completely different for typical white dwarf central stars ($\teff =
100000$\,K, $\log g \approx 7.0$). The higher temperature and lower gravity
causes large deviations between both types of model spectra. Therefore it is
necessary to use NLTE model atmospheres for the analysis of these stars. 

Levels and lines are included in NLTE up to $n=16$ for hydrogen, $n=32$ for
\ion{He}{ii}, and $n=9$ for \ion{He}{i}. \ion{He}{i} levels with $n=5$ to~9
were merged into one singlet and one triplet level for every principal quantum
number. The $n=5$ levels were split with respect to their angular quantum
number for the subsequent spectrum synthesis. Line blanketing by the Stark
broadened lines is taken into account consistently for hydrogen, \ion{He}{ii}
lines and the resonance series of \ion{He}{i}. The synthetic spectra are
computed with the extended VCS broadening tables (Vidal et al.\ \cite{VCS70})
provided by Sch\"oning \& Butler (\cite{SB89}; priv.\ comm.) and Lemke
(\cite{L97}) for hydrogen and \ion{He}{ii}. We used the \ion{He}{i} line
broadening tables computed by Barnard et al.\ (\cite{BCS74}), Shamey
(\cite{S64}) and Barnard et al.\ (\cite{BCS69}). Profiles of the isolated
\ion{He}{i} lines 5016\,\AA\ and 5876\,\AA\ were computed according the
prescription of Griem (\cite{G74}). 
 
Pressure dissolution of the higher levels is described by  the Hummer \&
Mihalas (\cite{HM88}) occupation probability formalism following the NLTE
implementation by Hubeny et al.\ (\cite{HHL94}). In contrast to our analysis of
EUV selected DA white dwarfs (Napiwotzki et al.\ \cite{NGS99}) we did {\em not}
increase the critical ionizing field $\beta_{\mathrm{crit}}$ adopted to
calculate the occupation probability by a factor two as proposed by Bergeron
(\cite{B93}). The motivation of Bergeron was not a flaw in the Hummer \&
Mihalas (\cite{HM88}) formalism, but a compensation for the inadequacy of the
standard Stark broadening theory when line wings overlap. This improved the
agreement with experimental measurements and the fits to DA white dwarfs in the
Bergeron et al.\ (\cite{BSL92}) sample. Finley et al.\ (\cite{FKB97}) reported
also better fit quality for hotter EUV selected white dwarfs. However, the
central stars analysed here have even higher temperatures, and their gravities
are lower by typical one or two dex in comparison to the stars discussed in
Bergeron (\cite{B93}) and Finley et al.\ (\cite{FKB97}). It is by no means
clear if the correction factor is a valid approximation in the case of our
CSPN. Thus we preferred to leave the original value of the critical ionizing
field $\beta_{\mathrm{crit}}$ unchanged. From our experience with the EUV
selected DA white dwarfs analysed by Napiwotzki et al.\ (\cite{NGS99}) we
estimate offsets of the order 0.1\,dex in gravity and 2\% for $\teff$ when
compared to an analysis, which applies this correction factor. 

Our NLTE model grids have to cover wide ranges in temperature
($30000\,\mathrm{K} \le \teff \le 300000$\,K), gravity ($4.75 \le \log g \le
8.75$), and helium abundance (from pure hydrogen up to $\nhe/\nh = 3$). Thus,
to keep computational effort within reasonable limits, we didn't compute every
model in this 3D-mesh, but only the needed ones. 

Due to the common lack of other temperature indicators both, effective
temperature and gravity, must be determined from the Balmer lines. For DA white
dwarfs this was successfully done by simultaneous line profile fitting of all
available Balmer lines (e.g.\ Bergeron et al.\ \cite{BSL92}; Napiwotzki et al.\
\cite{NGS99}). Generally the observed line profiles are well reproduced by
model spectra with the optimum parameters.

\begin{figure}
\epsfxsize=5.4cm
\epsffile{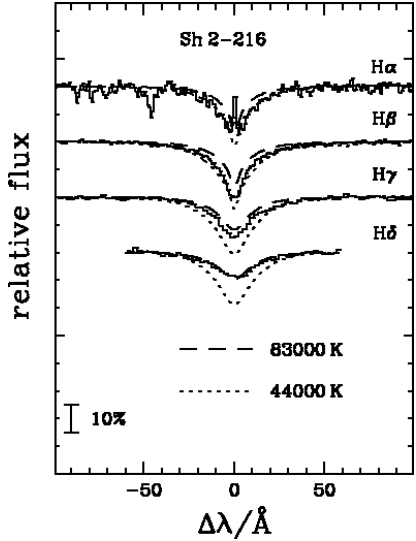}
\caption{Model spectra calculated for two different temperatures are compared
to the Balmer lines of \object{LS\,V+46\,21}, the central star of
\object{Sh\,2-216}} 
\label{f:s216problem}
\end{figure}

However, this method failed for nearly all hydrogen-rich central stars of old
PNe (Napiwotzki \cite{N92}; Napiwotzki \& Rauch \cite{NR94}): no consistent fit
to the Balmer lines (\halpha\ to \hdelta\ were used) was possible. A strong
trend is present: fitting of higher Balmer lines yields higher temperatures.
This effect is demonstrated for the spectrum of the white dwarf central star
\object{Sh\,2-216} in Fig.~\ref{f:s216problem}. While a high $\teff$ model
(83000\,K) fits the \hdelta\ line, only 44000\,K are needed to fit the \halpha\
line. This difference causes serious problems for the interpretation of the
evolutionary status of white dwarf central stars. The previous investigation
carried out by M\'endez et al.\ (\cite{MKH88a}) restricted the analysis to one
Balmer (\hgamma) line only and thus failed to recognize this discrepancy. 

Napiwotzki (\cite{N93a}) argued that the high temperature derived from the
highest Balmer line available (\hdelta\ or \hepsilon\ respectively) is closest
to the real temperature of the white dwarf central stars. Important evidence
came from the analysis of the sdO star \object{BD$+28^\circ 4211$}. Its
spectral appearance is very similar to central stars of old PNe. As can be seen
from Fig.~3 in Napiwotzki (\cite{N93a}) the Balmer line problem is of similar
strength as it is in the central star of \object{Sh\,2-216}
(Fig.~\ref{f:s216problem}). Since \object{BD$+28^\circ 4211$} is relatively
bright ($V=10.5$) we were able to obtain a high-resolution spectrum and
detected the 5876\,\AA\ line of neutral helium (Fig.~4 of Napiwotzki
\cite{N93a}). \ion{He}{i} lines are very sensitive temperature indicators in
this parameter range. We derived a best fit with $\teff = 82000$\,K in
excellent agreement with the temperature determination from the \hepsilon\ line
(85000\,K). 
 
Napiwotzki \& Sch\"onberner (\cite{NS93}) and Napiwotzki \& Rauch (\cite{NR94})
discussed possible reasons of the Balmer line problem: wind effects, magnetic
fields, pressure ionization and line quenching, deficits in the line broadening
theory, and modifications of the atmospheric structure due to line blanketing
of heavy elements. The authors concluded that none of these effects is
sufficient to explain the Balmer line problem. Kub\'at (\cite{K95}) found a
moderate sphericity effect on \halpha\ and, to a lesser extent, on \hbeta, too,
for pre-white dwarfs with $\log g \approx 6$. These effects are considerably
weaker for larger gravity and too small to account for the Balmer line problem.

Finally, Werner (\cite{W96}) presented the likely solution of this problem. He
demonstrated that the inclusion of Stark broadening for C, N, and O lines can
have a strong influence on the atmospheric structure of very hot hydrogen-rich
stars. It is likely that the effect on the emergent spectrum is pronounced
enough to solve the Balmer line problem. An application of his model
atmospheres on \object{BD$+28^\circ 4211$} gave good agreement with the
observed spectrum, if the parameters of Napiwotzki (\cite{N93a}) were used.
Werner's results confirm Napiwotzki's (\cite{N93a}) sophisticated guess that
the temperatures derived from the highest Balmer lines are the most reliable,
close to the ``real'' temperature of the central stars. The reason for this
behavior is evident in Fig.~1 of Werner (\cite{W96}), which displays the
temperature structure of model atmospheres computed with different treatment of
line opacity. The inclusion of Stark broadened CNO lines causes a strong
modification of the temperature structure in the atmospheric layers, where
\halpha\ and \hbeta\ are formed, while the change in the deeper layers, where
\hdelta\ and \hepsilon\ are formed is negligible. 

However, to apply this procedure the abundances of C, N, and O need to be known
which is not the case for our sample. In addition the computation of a NLTE
model grid accounting for the influence of C, N, and O in the above described
manner requires very large amounts of computer time, hence we analysed the CSPN
with our atmospheres containing only H and He. Since the Balmer line fits
result in the same gravity for every Balmer line (within the error limits) we
used the following recipe: $g$ was computed from the average of all Balmer
lines and $\teff$ was derived from the fit of \hdelta\ (and \hepsilon\ if
available) with $g$ fixed at the average value. The temperatures derived in our
early analyses (Napiwotzki \& Sch\"onberner \cite{NS91b} and partly in
Napiwotzki \cite{N93b}) are often too low, because we could not use the
\hdelta\ line at that time due to reduction problems. All our new analyses now
include this line and the discrepant cases discussed in Pottasch (\cite{P96})
with Zanstra temperatures much higher than our analysis results are now
resolved. 

The line fits were performed with the least-square algorithm described in
Bergeron et al.\ (\cite{BSL92}). The observed and theoretical Balmer line
profiles are normalized to a linear continuum in a consistent manner. 
Wavelength shifts are determined with a cross-correlation method and applied
consistently to each complete spectrum.  The synthetic spectra are convolved to
the observational resolution with a Gaussian and interpolated to the actual
parameters with bicubic splines, and interpolated to the observed wavelength
scale. 

The atmospheric parameters are then determined by minimizing the $\chi^2$ value
by means of a Levenberg-Marquardt steepest descent algorithm (Press et al.\
\cite{PTV92}). Finally, an estimate of the internal errors can be derived from
the covariance matrix. In contrast to Bergeron et al.\ (\cite{BSL92}), we
estimate the noise of the spectra ($\sigma$) used for the $\chi^2$ fit from the
neighboring continuum of each line. The S/N is adopted to be constant
throughout the line. Line cores, which show some remaining contamination from
the nebula, were excluded from the fits. 

In our investigation of EUV selected DA white dwarfs (Napiwotzki et al.\
\cite{NGS99}) we showed that the internal errors are usually much smaller than
the real observational scatter as inferred from a comparisons of independent
analysis. We used the data collection of Napiwotzki et al.\ (\cite{NGS99}) to
derive a more realistic estimate of observational errors. For this purpose we
binned the data into 5000\,K intervals, determined the scatter, and performed a
simple linear fit with $\log \teff$ as independent variable. The results are 

\begin{eqnarray}
\frac{\sigma(\teff)}{\teff} &= &-0.0995 +0.0273\cdot \log \teff \\
\sigma(\log g) 		&= &-0.8673 +0.2130\cdot \log \teff
\end{eqnarray}

A moderate extrapolation is necessary to apply these fits to our CSPN analyses.
We took a conservative approach and assumed that these $\sigma$ values
correspond completely to the external errors and added them quadratically to
the internal errors derived with the fit procedure. 

\section{Results and evolutionary status of the central stars}

The results of our analysis are summarized in Table~\ref{t:cpn}. The fits to
the Balmer lines and helium lines are displayed in Figs.~\ref{f:daogitter1}
to~\ref{f:coolgitter}. Due to the Balmer line problem discussed above we
derived usually different temperatures from the individual Balmer lines. The
individual values are indicated in the plots. The spectra are grouped according
to their spectral appearance. Within a plot the spectra are displayed in order
of increasing galactic longitude. 

\begin{table*}
\caption{Parameters of the hydrogen-rich central stars. $\teff$ was derived
from the fit of \hdelta\ (+\hepsilon) as described in the text. The CSPN are
denoted by their PN\,G designation and by their common names. Masses were
derived from the comparison with the evolutionary tracks in Fig.~\ref{f:HR}} 
\label{t:cpn}
\begin{center}
\begin{tabular}{ll|r@{$\pm$}lr@{$\pm$}lr@{.}l@{$\pm$}lr@{$\pm$}l|rrr}
PN\,G	&Name	&\multicolumn{2}{c}{$\teff$}	
	&\multicolumn{2}{c}{$\log g$}
	&\multicolumn{3}{c}{$\log \nhe/\nh$}	
	&\multicolumn{2}{c}{$M$}
	&\multicolumn{1}{|c}{$d_{\mathrm{NLTE}}$}	
	&\multicolumn{1}{c}{$R$}	
	&\multicolumn{1}{c}{$t_{\mathrm{kin}}$}	\\
	&	&\multicolumn{2}{c}{(K)}	
	&\multicolumn{2}{c}{$\mathrm{(cm\,s^{-1})}$}
	&\multicolumn{3}{c}{ }
	&\multicolumn{2}{c}{$(M_\odot)$}
	&\multicolumn{1}{|c}{(pc)}	
	&\multicolumn{1}{c}{(pc)}	
	&\multicolumn{1}{c}{$(10^3$yrs)}\\ 	\hline
\object{025.4$-$04.7}	&\object{IC\,1295}	&90100&6200	&6.66&0.30
				&$-$1&31&0.12	&0.51&0.04	
				&715	&0.16	&7.6\\
\object{027.6+16.9}	&\object{DeHt\,2}	&117000&6300	&5.64&0.22
				&$-$0&79&0.12
				&\multicolumn{2}{l|}{0.64$^{+0.13}_{-0.06}$}
				&2382	&0.54	&27\\
\object{030.6+06.2}	&\object{Sh\,2-68}	&95800&9300	&6.78&0.32
				&$-1$&02&0.21	&0.55&0.03	
				&1054	&1.02	&200\\
\object{034.1$-$10.5}	&\object{HDW\,11}	&68100&9400	&6.38&0.31	
				&$-$1&23&0.24	&0.39&0.03	
				&1176	&0.13	&6.6\\
\object{036.0+17.6}	&\object{A\,43}		&116900&5500	&5.51&0.22	
				&$-$0&14&0.22	
				&\multicolumn{2}{l|}{0.68$^{+0.13}_{-0.08}$}
				&2649	&0.51	&25	\\
\object{036.1$-$57.1}	&\object{NGC\,7293}	&103600&5500	&7.00&0.22	
				&$-$1&43&0.15	&0.57&0.02 
				&291	&0.69	&28	\\
\object{047.0+42.4}	&\object{A\,39}		&117000&11000	&6.28&0.22	
				&$-$0&85&0.10	&0.57&0.02	
				&1931	&0.81	&22	\\
\object{060.8$-$03.6}	&\object{NGC\,6853}	&108600&6800	&6.72&0.23
				&$-1$&12&0.09	&0.56&0.01	
				&436	&0.42	&13.4	\\
\object{063.1+13.9}	&\object{NGC\,6720}	&101200&4600	&6.88&0.26	
				&$-$1&14&0.09	&0.56&0.02	
				&1088	&0.20	&2.6	\\
\object{066.7$-$28.2}	&\object{NGC\,7094}	&125900&7700	&5.45&0.23	
				&$-$0&04&0.14
				&\multicolumn{2}{l|}{0.87$^{+0.20}_{-0.23}$}
				&2246	&0.51	&11.1	\\
\object{072.7$-$17.1}	&\object{A\,74}		&108000&15000	&6.82&0.27
				&$-1$&94&0.25	&0.56&0.03	
				&1673	&4.52	&170	\\
\object{077.6+14.7}	&\object{A\,61}		&88200&7900	&7.10&0.37
				&$-$1&58&0.18	&0.55&0.05 
				&1380	&0.67	&22	\\
\object{111.0+11.6}	&\object{DeHt\,5}	&76500&5800	&6.65&0.19	
				&\multicolumn{3}{c}{$<$$-$2.69}	&0.44&0.04
				&512	&0.66	&129	\\
\object{120.3+18.3}	&\object{Sh\,2-174}	&69100&3000	&6.70&0.18	
				&$-2$&55&0.13	&0.43&0.03	
				&556	&0.81	&40	\\
\object{124.0+10.7}	&\object{EGB\,1}	&147000&25000	&7.34&0.31	
				&\multicolumn{3}{c}{$<$$-$1.66}	&0.65&0.05
				&653	&0.41	&20	\\
\object{125.9$-$47.0}	&\object{PHL\,932}	&35000&900	&5.93&0.12
				&$-$1&53&0.05	&0.28&0.01
				&235	&0.16	&7.7	\\
\object{128.0$-$04.1}	&\object{Sh\,2-188}	&102000&32000	&6.82&0.60
				&$-$1&27&0.64	&0.56&0.07	
				&965	&0.88	&22	\\
\object{148.4+57.0}	&\object{NGC\,3587}	&93900&5600	&6.94&0.31
				&$-1$&07&0.13	&0.55&0.03
				&1269	&0.52	&12.8	\\
\object{149.4$-$09.2}	&\object{HDW\,3}	&125000&28000	&6.75&0.32
				&$-$0&87&0.37	&0.58&0.03	
				&1451	&1.90	&93	\\
\object{156.3+12.5}	&\object{HDW\,4}	&47300&1700	&7.93&0.16	
				&\multicolumn{3}{c}{$<$$-$3.59}	&0.64&0.07
				&246	&0.063	&3.1	\\
\object{156.9$-$13.3}	&\object{HaWe\,5}	&38100&1500	&7.58&0.20	
				&\multicolumn{3}{c}{$<$$-$2.21}	&0.51&0.04
				&420	&0.035	&1.7	\\
\object{158.5+00.7}	&\object{Sh\,2-216}	&83200&3300	&6.74&0.19	
				&$-$1&95&0.06	&0.49&0.03	
				&185	&2.69	&660	\\
\object{158.9+17.8}	&\object{PuWe\,1}	&93900&6200	&7.09&0.24	
				&$-$1&70&0.20	&0.56&0.03	
				&695	&2.02	&73	\\
\object{197.4$-$06.4}	&\object{WeDe\,1}	&141000&19000	&7.53&0.32	
				&\multicolumn{3}{c}{$<$$-1$.70}	&0.68&0.07
				&968	&2.17	&133	\\
\object{204.1+04.7}	&\object{K\,2-2}	&67000&11000	&6.09&0.24
				&$-$1&55&0.15	&0.38&0.04	
				&628	&0.66	&65	\\
\object{215.5$-$30.8}	&\object{A\,7}		&99000&18000	&7.03&0.43	
				&$-$1&49&0.37	&0.57&0.05	
				&705	&1.30	&63	\\
\object{219.1+31.2}	&\object{A\,31}		&84700&4700	&6.63&0.30
				&$-$1&53&0.13	&0.48&0.04 
				&988	&2.32	&65	\\ \hline
		&\object{HZ\,34}		&90800&3900	&6.60&0.20
				&$-$1&59&0.09	&0.51&0.03	
				&	&	&	\\
\end{tabular}
\end{center}
\end{table*}

\begin{figure*}
\epsfxsize=17cm
\epsffile{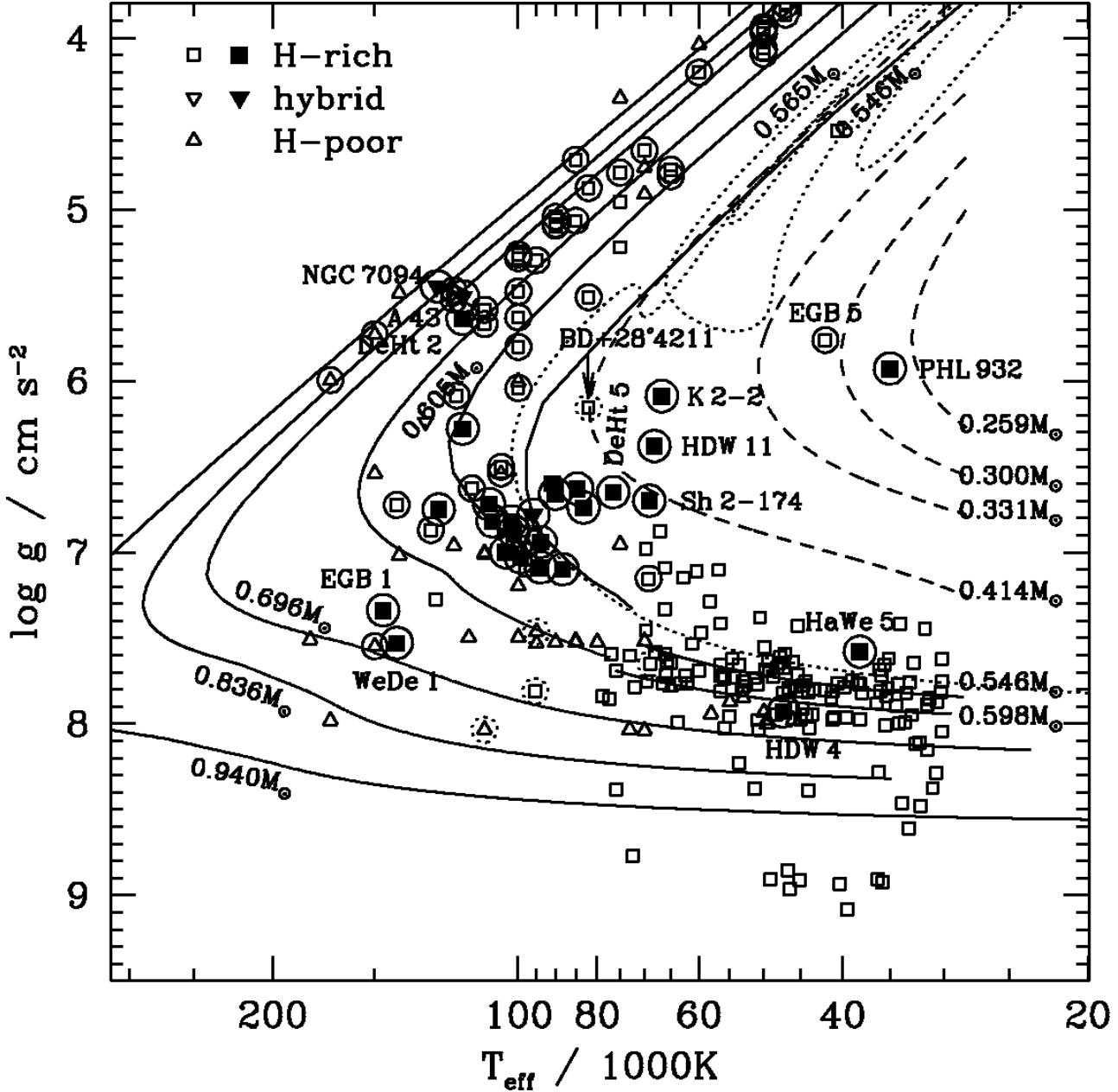}
\caption[]{
$\teff$-$g$ diagram with the results of the new analyses plotted as filled
symbols. These data are supplemented by results taken from literature (open
symbols). Hydrogen-rich objects are marked by squares, hydrogen-poor by
triangles. Central stars are encircled, a dashed circle is used for suspected
PNe. Since the parameters from literature are sometimes rounded to identical
values the positions of these stars are shifted by small random values.
Post-AGB evolutionary tracks are drawn as solid lines, labeled with the mass of
the remnant. Dashed lines indicate the post-RGB tracks discussed in the text.
The $0.524M_\odot$ post-EAGB track is drawn as dotted line 
\label{f:HR}}
\end{figure*}

\begin{figure}
\epsfxsize=8.5cm
\epsffile[18 58 590 760]{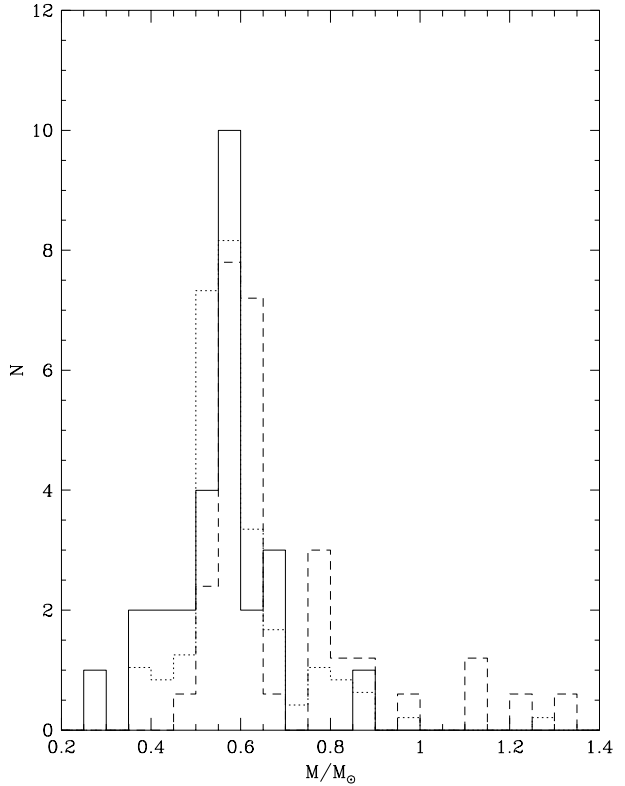}
\caption[]{Mass distribution of the central stars of old PN (solid
histogram) compared to two mass distributions obtained from analyses
of hot DA white dwarfs (Bergeron et al.\ \cite{BSL92}, dotted histogram;
Napiwotzki et al.\ \cite{NGS99}, dashed histogram)}
\label{f:massdist}
\end{figure}

A $\teff$-$g$ diagram with the results is given in Fig.~\ref{f:HR}. For
comparison evolutionary tracks of post-AGB stars computed by Bl\"ocker
(\cite{B95}), Sch\"onberner (\cite{S83}), Koester \& Sch\"onberner
(\cite{KS86}) labeled with the remnant mass are plotted as solid lines. The
dashed low mass tracks are taken from the computations of Driebe et al.\
(\cite{DSB98}). They represent the evolution of stars whose hydrogen-rich
envelope was stripped away by a companion during the (first) red giant branch
phase (Kippenhahn et al.\ \cite{KKW67}; Iben \& Tutukov \cite{IT86}). The
remaining mass is too low for the ignition of helium burning and the star will
eventually become a low mass white dwarf with a helium core. In the remainder
of this paper we will term this scenario ``post-RGB evolution'' for the sake of
simplicity. 

The dotted track describes the evolution of a $0.524M_\odot$ post-early AGB
(post-EAGB) track from Bl\"ocker (\cite{B95}). This star suffers from two late
helium flashes, which are responsible for the extended loops, and eventually
enters the white dwarf domain on a track very similar to that of the
$0.546M_\odot$ model. Helium burning stars with significantly lower masses will
never enter the (early) AGB and thus never produce a PN (AGB manqu\'e
evolution; cf.\ Dorman et al.\ \cite{DRO93}). 

The new analyses are plotted as filled symbols. Our new results are
supplemented by data taken from the literature (open symbols) for hydrogen-rich
CSPN and DA/DAO white dwarfs. Hydrogen-poor objects are represented by PG\,1159
stars, DO white dwarfs, and O(He) post-AGB stars. Stellar parameters and
references are provided in Table~\ref{t:lit} and Table~\ref{t:ref} in the
Appendix. The symbols of stars with a known PN are encircled. Although analyses
of hydrogen-poor Wolf-Rayet CSPN exist (Koesterke \& Hamann \cite{KH97}), which
indicate that these stars are counterparts of the luminous hydrogen-rich CSPN
analyzed by M\'endez et al.\ (\cite{MKH88a}) there is no simple way to place
them into our $\teff$-$g$ diagram, because a $g$ determination is not possible
without further assumptions. 

The central stars of old PNe fill nicely the reported ``gap'' of the
hydrogen-rich sequence. A continuous hydrogen-rich sequence from the central
star to the white dwarf region is revealed, and therefore evolutionary
scenarios claiming a hydrogen-poor stage of all (or most) pre-white dwarfs
(Fontaine \& Wesemael \cite{FW87}) can be ruled out. The gap in the
hydrogen-rich sequence can be traced back to selection effects. Some of the
hottest hydrogen-rich white dwarfs in the Palomar-Green survey (Green et al.\
\cite{GSL86}), the most important source of faint blue stars until recently,
were originally misclassified as sdO or sdB stars (cf.\ Jordan et al.\
\cite{JHW91}, Liebert \& Bergeron \cite{LB95}). It is much easier to pick out
the peculiar hydrogen-poor PG\,1159 stars and DO white dwarfs. In addition,
previous analysis of very hot DA or DAO white dwarfs tended to produce too low
temperatures. This is probably caused by the neglect of NLTE effects and the
Balmer line problem in these stars (cf.\ the discussion in Liebert \& Bergeron
\cite{LB95}). 

We determined masses of the central stars by interpolating in the evolutionary
tracks in Fig.~\ref{f:HR}. The values are provided in the sixth column of
Table~\ref{t:cpn}. Error limits were estimated by propagating the errors of
temperature and gravity. Additional errors may result from the adopted tracks.
This is not a major problem for stars lying on the post-AGB tracks. However,
the mass determination of stars, which cannot be explained by standard post-AGB
evolution, might be wrong, if the star didn't evolve according to the adopted
scenario but has a completely different history. The individual cases are
discussed below. 

The resulting mass distribution (binned over $0.05M_\odot$) is shown in
Fig.~\ref{f:massdist}. For comparison we show mass distributions derived for
hot DA white dwarfs (Bergeron et al.\ \cite{BSL92}; Napiwotzki et al.\
\cite{NGS99}) scaled down to the number of analyzed CSPN. The Bergeron et al.\
mass distribution was redetermined with the Bl\"ocker (\cite{B95}) and Driebe
et al.\ (\cite{DSB98}) evolutionary tracks. The Napiwotzki et al.\
(\cite{NGS99}) analysis already applied these tracks. The Bergeron et al.\
sample contains mainly relatively cool DA white dwarfs in the temperature
interval $15000\,\mathrm{K} \le \teff \le 30000$\,K, while the Napiwotzki et
al.\ sample is EUV selected and covers $25000\,\mathrm{K} \le \teff \le
55000$\,K. The CSPN mass distribution peaks in the $0.55\ldots 0.60$ interval,
similar to the white dwarf distributions. Note, however, that the low mass tail
is more pronounced than in the DA studies. Napiwotzki et al.\ described that
EUV selected samples do select {\em against} low mass white dwarfs, which
explain the deficiency in their study. It is likely that slowly evolving low
mass stars are {\it preferentially} detected in extended, low density nebulae.
However, we don't expect a strong selection against high mass CSPN in our
sample, because the evolution of massive central stars slows down and becomes
comparable to that of low mass stars, when low luminosities are reached.
Bl\"ocker (\cite{B95}) claims that high mass CSPN evolve even slower than
$0.6M_\odot$ remnants in this phase. However, this depends on the treatment of
AGB mass loss (cf.\ discussions in Bl\"ocker \cite{B95} and Vassiliadis \& Wood
\cite{VW94}). 

Since the mean might be heavily biased by few objects with extreme masses, it
is not very useful for the comparison of white dwarf mass distributions. Finley
et al.\ (\cite{FKB97}) proposed a Gaussian fit of the mass peak as a robust
estimate of the distribution peak. Due to small number statistics and/or the
high fraction of low mass stars the width of the best fitting Gauss curve is
relatively high: $0.14M_\odot$. This is twice the value typically found in the
DA investigations and casts some doubt on the fit quality. However, we redid
the fit with the width held fixed at $0.07M_\odot$ and computed a peak mass
only $0.001M_\odot$ higher. This convinced us that the derived peak mass of
$0.55M_\odot$ is reliable. The corresponding value for the Bergeron et al.\
(\cite{BSL92}) sample ($0.56M_\odot$) and for the Napiwotzki et al.\
(\cite{NGS99}) sample ($0.59M_\odot$) are in agreement. Stasi\'nska et al.\
(\cite{SGT97}) determined the mass distribution of CSPN from a distant
independent approach using the nebular \hbeta\ flux, the angular radius, the
expansion velocity, and the stellar $V$ magnitude. This method is based on a
simple model of the PN evolution (G\'orny et al.\ \cite{GST97}). Stasi\'nska et
al.\ find a narrow mass distribution with $\approx$80\% of all objects in the
range $0.55\ldots 0.65M_\odot$. The median of the sample amounts to
$0.60M_\odot$. However, this value is slightly dependent of the adopted nebula
mass in their PN model. Although this is a completely different approach it is
in good agreement with our results for the central stars of old PNe. However,
Stasi\'nska et al.\ find no central stars with masses below $0.55M_\odot$. This
might partly be explained by their selection criteria, which prefer brighter,
younger PNe. 

We have used the values of effective temperature, gravity, and mass given in
Table~\ref{t:cpn} and the apparent magnitudes given in Paper~III to calculate
the distances $d_{\mathrm{NLTE}}$ of the CSPN. The results are provided in
Table~\ref{t:cpn} together with the radius $R$ and the kinematical age
$t_{\mathrm{kin}}$ of the PNe computed from the nebular data compiled in
Paper~III. See Paper~V for more details. 

\subsection{The DAO stars}

\begin{figure*}
\epsfxsize=17.0cm
\epsffile{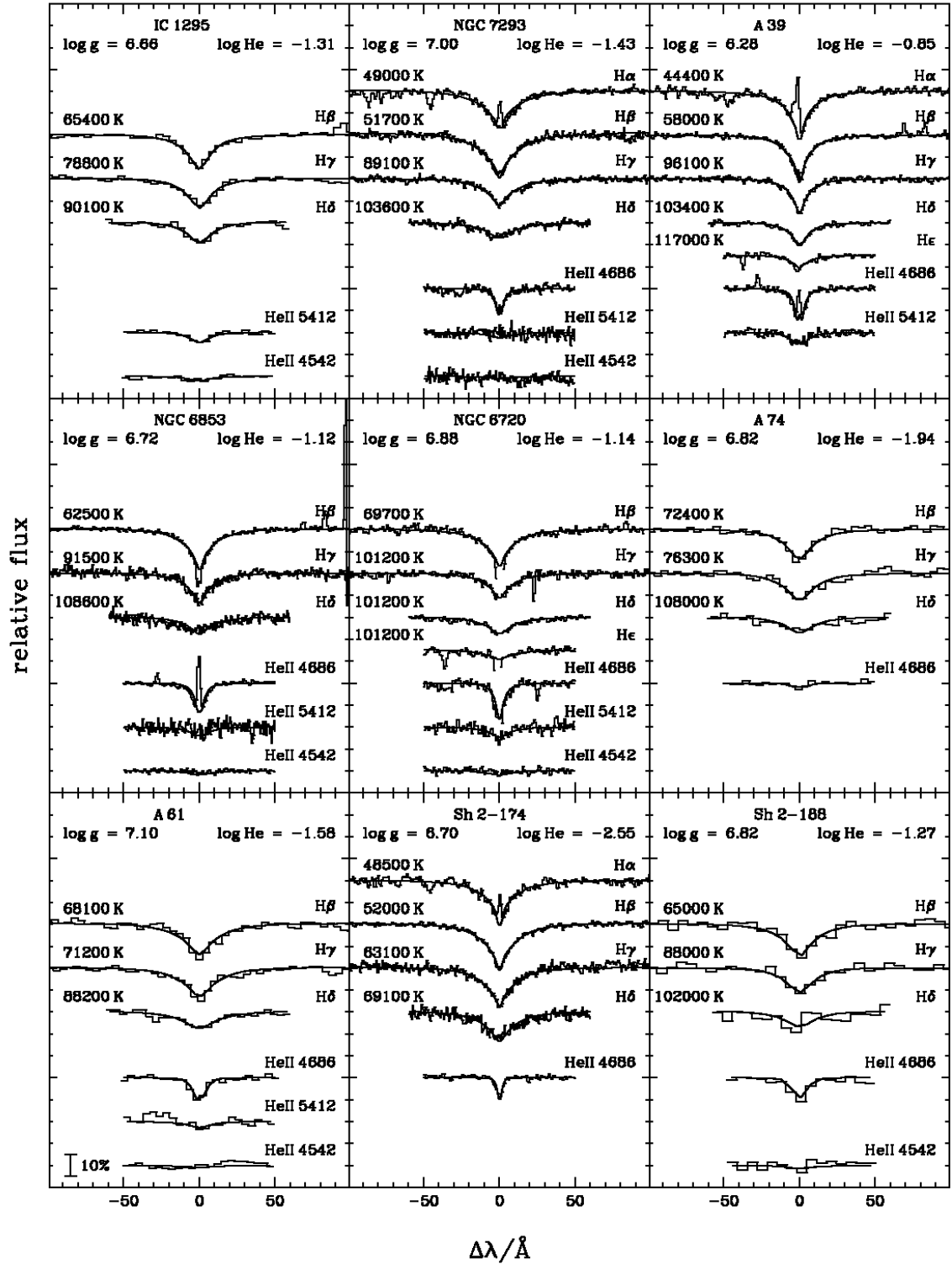}
\caption{Spectral fits of DAO central stars. The stars are ordered by
increasing galactic longitude}
\label{f:daogitter1}
\end{figure*}

\begin{figure*}
\epsfxsize=17.0cm
\epsffile{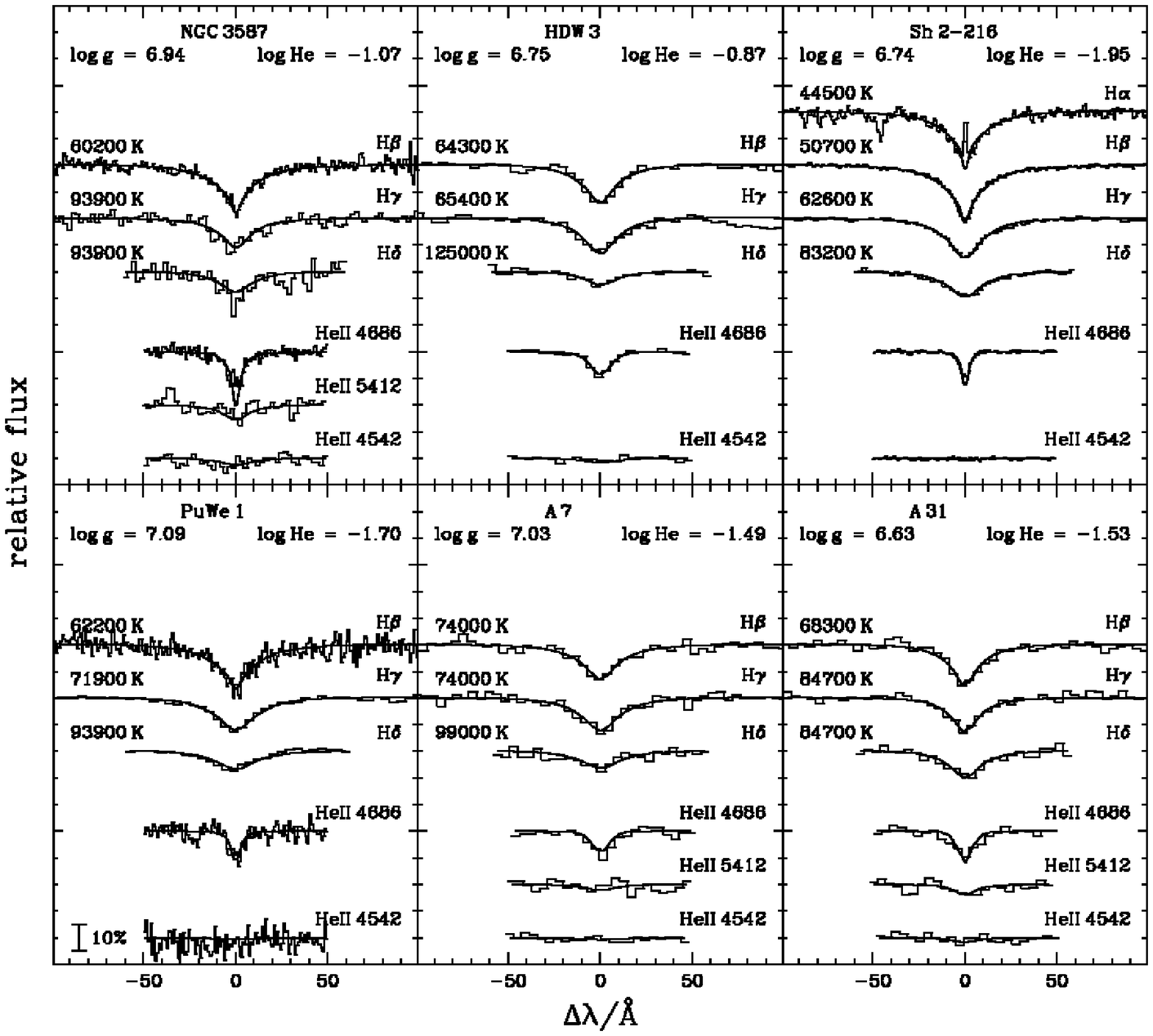}
\caption{Spectral fits of DAO central stars (cont.)}
\label{f:daogitter2}
\end{figure*}

Most central stars of old PNe cluster around $\teff \approx 100000$\,K and
$\log g \approx 7.0$. These are the DAO central stars displayed in
Figs.~\ref{f:daogitter1} and~\ref{f:daogitter2}. Among this group the central
stars of well known PNe like \object{NGC\,6720} (the Ring nebula),
\object{NGC\,7293} (the Helix nebula), and \object{NGC\,6853} (the Dumbbell
nebula) are found. Another member is the central star of \object{Sh\,2-216},
the closest known PN. According to the definition of the spectral type DAO the
spectra are dominated by the hydrogen Balmer lines, but the \ion{He}{ii}
4686\,\AA\ line and sometimes \ion{He}{ii} 4542\,\AA\ and 5412\,\AA\ are
present. Generally these DAO central stars compare well with low mass post-AGB
tracks. That's in contrast with the results Bergeron et al.\ (\cite{BWB94}) for
field DAO white dwarfs. These stars are cooler and are probably descendents of
the extreme horizontal branch (EHB), the so-called AGB-manqu\'e stars, which
stay hot during their post-HB evolution and never enter the AGB. Thus we
conclude that DAO stars can result from both evolutionary paths. In
Sect.~\ref{s:helium} we will show that, according to the observational
evidence, the DAO phase is simply a transition stage passed by stars entering
the DA sequence. 

Note that our re-analysis of \object{HZ\,34} yields a temperature 10000\,K
higher than the value given by Bergeron et al.\ (\cite{BWB94}). The new
parameters are consistent with a post-AGB nature. Actually this star has
parameters very similar to other DAO central stars with a nebula. Therefore it
is tempting to speculate about an as yet undetected PN around this star.
However, a deep survey for nebula emission performed by Tweedy \& Kwitter
(\cite{TK94}) ended up with a negative result for \object{HZ\,34}. Another star
with parameters similar to the DAO central stars is the well-known sdO star
\object{BD$+28^\circ 4211$}. The analysis presented in Napiwotzki (\cite{N93a})
yielded $\teff = 82000$\,K, $\log g = 6.2$, and $\nhe/\nh = 0.1$. These
parameters place \object{BD$+28^\circ 4211$} in a region of the HR diagram,
which is consistent with a star either close to the lower mass limit of
post-AGB evolution or with post-RGB evolution. Interestingly, Zanin \&
Weinberger (\cite{ZW97}) detected recently a very large, faint emission region
with an angular diameter of $5^\circ$ centered on \object{BD$+28^\circ 4211$},
which may possibly be the PN of this star. However, more observational data is
needed before one can decide whether this is a chance alignment or really the
PN of \object{BD$+28^\circ 4211$}. 

\begin{figure}
\epsfxsize=5.4cm
\epsffile{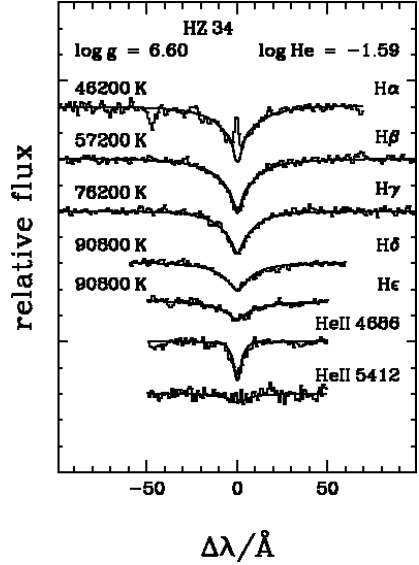}
\caption{Our reanalysis of the field DAO white dwarf \object{HZ\,34} from the 
Bergeron et al.\ (\cite{BWB94}) sample}
\label{f:hz34gitter}
\end{figure}

The coolest DAO white dwarf found in our sample is \object{GD\,561}, the
central star of \object{Sh\,2-174}. It has an effective temperature of only
69000\,K and cannot be explained by standard post-AGB evolution (cf.\
Fig.~\ref{f:HR}). The status of \object{GD\,561} was already discussed in
Tweedy \& Napiwotzki (\cite{TN94}). The existence of a PN provides us with
constraints on possible evolutionary scenarios. Bergeron et al.\ (\cite{BWB94})
proposed an origin from the extreme horizontal branch (EHB). The parameters of
\object{GD\,561} can be reproduced by post-EHB evolution (e.g.\ Dorman et al.
\cite{DRO93}). However the PN remains unexplained. Since the horizontal branch
evolution lasts $\approx$$10^{8}$ years, any nebula produced on the RGB would
be dispersed in the meantime. The mass of the remaining hydrogen-layer
($<0.01M_\odot$) of EHB stars is much too low to produce a relevant nebula at
the end of the horizontal branch stage. Obviously this region of the HR diagram
is populated by stars which are not of a post-EHB origin as well. 

\object{GD\,561} is most likely the outcome of close binary evolution. During 
the giant stage (on the AGB or RGB) of the central star precursor, it can fill
its Roche lobe or even enclose the companion in its envelope (common envelope
evolution). Especially the latter causes heavy mass loss and a modification of
the stellar evolution. The result is a hot star inside an expanding shell,
which can mimic a normal PN (see e.g.\ the review of Livio \cite{L93}). Tweedy
\& Napiwotzki (\cite{TN94}) showed that the parameters of \object{GD\,561} can
be explained by the track of a $0.296M_\odot$ star, which lost its envelope
during a common envelope event on the RGB (Iben \& Tutukov \cite{IT86}).
However, the evolutionary age ($4\cdot 10^6$\,yrs) predicted by these
calculations is uncomfortably large compared with the kinematical age of
\object{Sh\,2-174} (40000\,yrs; Table~\ref{t:cpn}). The new post-RGB tracks of
Driebe et al.\ (\cite{DSB98}) provide a much better agreement: the
$0.414M_\odot$ track reaches the position of \object{GD\,561} within
100000\,years after the RGB is left. Note that the PN age estimate is rather
uncertain because the expansion velocity has not been measured. A photometric
search for cool companions in the infrared J, H, K bands (Napiwotzki
\cite{N95}) found a cool companion of \object{GD\,561} separated by $\approx
4''$. Furthermore, the preliminary analysis revealed that the central star
itself has an infrared excess of $0\mag 5$ in K indicating binarity. The system
might in fact be a triple one. 

\subsection{The peculiar central stars \object{K\,2-2}, \object{HDW\,11}, and 
\object{PHL\,932}}

\begin{figure}
\epsfxsize=5.4cm
\epsffile{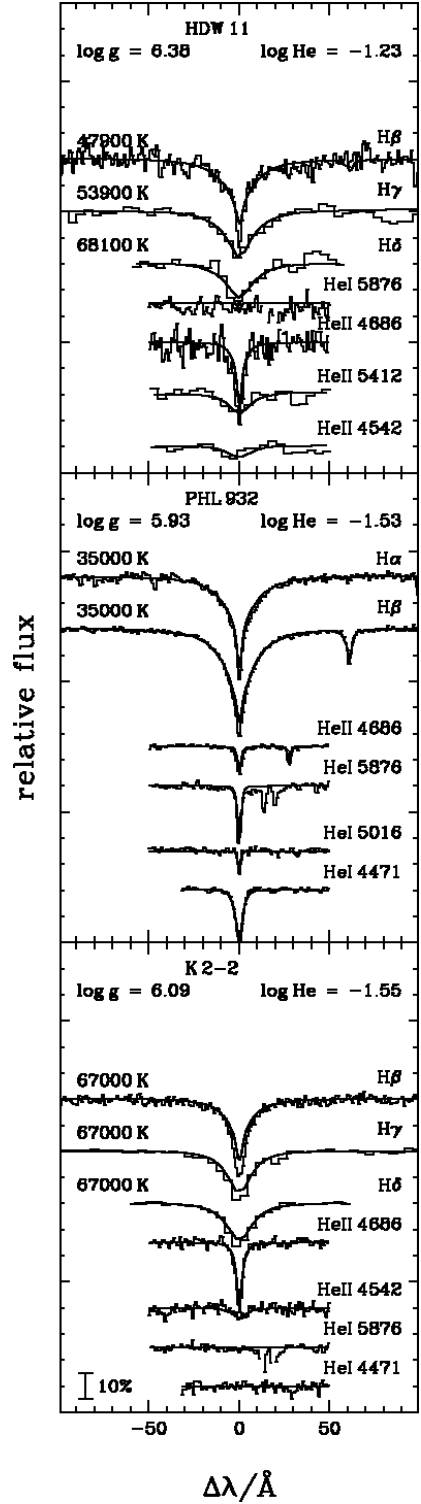}
\caption{Spectral fits of the peculiar central stars \object{HDW\,11},
\object{PHL\,932}, and \object{K\,2-2}}
\label{f:coolgitter}
\end{figure}

It appears that the CSPN \object{K\,2-2} and \object{HDW\,11}
(Fig.~\ref{f:coolgitter}) are lower gravity counterparts of
\object{GD\,561}/\object{Sh\,2-174}. Their positions in the HR diagram
(Fig.~\ref{f:HR}) fit into the post-RGB scenario. Judged from the comparison
with the Driebe et al.\ (\cite{DSB98}) tracks their masses are slightly lower
($0.39M_\odot$ and $0.38M_\odot$ for \object{HDW\,11} and \object{K\,2-2},
respectively) and their theoretical post-RGB ages lie in the range
$100000\ldots 200000$\,yrs. While the kinematical age of \object{K\,2-2}
(65000\,yrs) is consistent with that prediction, the age of \object{HDW\,11}
appears to be too low (6600\,years). Note that, again, no expansion velocity
has been measured for \object{HDW\,11} and thus the age was calculated by using
the canonical value of 20\,km/s. If the expansion is slower the age of
\object{HDW\,11} might be higher. 

While the age determination of \object{HDW\,11} might be in marginal agreement
with the post-RGB scenario, this is certainly not true for \object{PHL\,932}.
From Fig.~\ref{f:HR} we read of a mass of $0.28M_\odot$. The age corresponding
to the position of \object{PHL\,932} is $3\cdot 10^6$\,yrs, the PN age amounts
to 7700\,yrs only. Although both age determinations are subject to severe
uncertainties the discrepancy is so large that this scenario can be ruled out.
\object{PHL\,932} might be the result of a common envelope event of a star with
a degenerate CO core as calculated by Iben \& Tutukov (\cite{IT85}; see Iben \&
Livio \cite{IL93} for a review on common envelope evolution). M\'endez et al.\
(\cite{MGH88b}) speculated that both stars of the binary might have merged
during the common envelope stage. Note that our mass estimate is only valid if
\object{PHL\,932} is a post-RGB star. 

\subsection{The hot, high gravity central stars of \object{EGB\,1} and \object{WeDe\,1}}

\begin{figure}
\epsfxsize=5.4cm
\epsffile{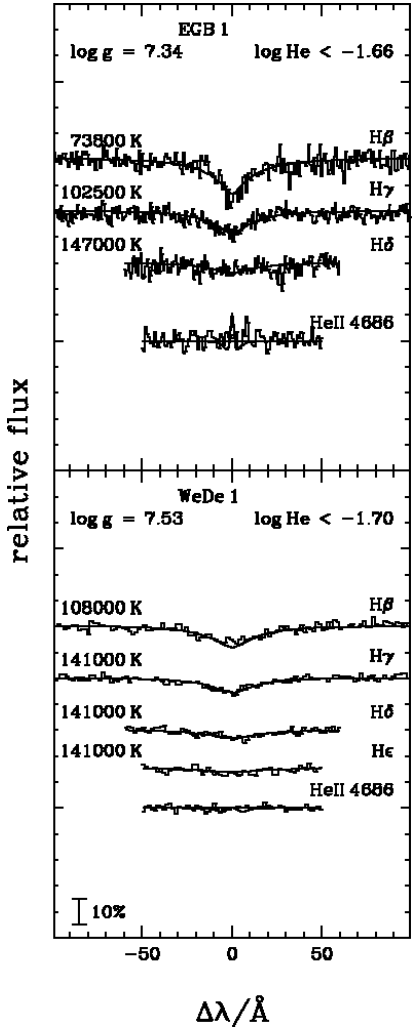}
\caption{The hot, high gravity central stars \object{EGB\,1} and \object{WeDe\,1}. 
The synthetic spectra of the \ion{He}{ii} 4686\,\AA\ line
are computed for the $3\sigma$ limit of the helium abundance}
\label{f:highlogg}
\end{figure}

Two high gravity CSPN are distinctly hotter and more massive than typical DAO
central stars: \object{EGB\,1} ($0.65M_\odot$) and \object{WeDe\,1}
($0.68M_\odot$). Their spectra don't show the \ion{He}{ii} 4686\,\AA\ line
indicating that their spectral type is DA. However, due to the high temperature
and gravity of these objects the upper limits, which can be derived from our
optical spectra are not very stringent: $\nhe/\nh < 0.02$ for both stars. Thus
the absence of helium lines can already be explained by a modest depletion. 

\subsection{DA central stars}

\begin{figure}
\epsfxsize=5.4cm
\epsffile{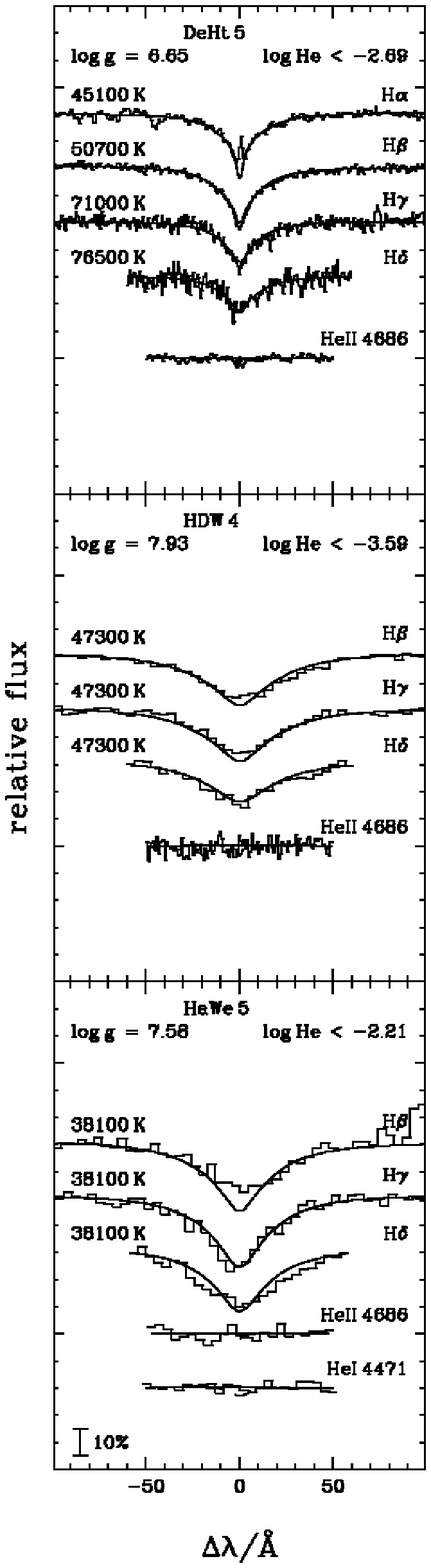}
\caption{DA central stars. The synthetic spectra of the helium lines
are computed for the $3\sigma$ limit of the helium abundance}
\label{f:dagitter}
\end{figure}

Three cooler DA white dwarfs are present in our sample, too
(Fig.~\ref{f:dagitter}). \object{DeHt\,5} has similar parameters as the DAO
white dwarf \object{GD\,561}/\object{Sh\,2-174} although it is slightly hotter
($\teff = 76500$\,K vs.\ 69100\,K). This places \object{DeHt\,5} closer to the
post-AGB tracks (Fig.~\ref{f:HR}), but a non post-AGB nature of
\object{DeHt\,5} is more likely. Several scenarios are possible (see discussion
of \object{GD\,561} above), but \object{DeHt\,5} compares well with the
post-RGB evolutionary tracks of Driebe et al.\ (\cite{DSB98}). The kinematical
age of the PN (129000\,years; Table~\ref{t:cpn}) is in good agreement with the
post-RGB age ($\approx$100000\,years) estimated from the $0.414M_\odot$ track. 

The DA white dwarfs \object{HDW\,4} and \object{HaWe\,5} are more mysterious.
Their spectral appearance (Fig.~\ref{f:dagitter}) and stellar parameters
($\teff = 47300$\,K, $\log g = 7.93$ and $\teff = 38100$\,K, $\log g = 7.58$)
resemble ordinary hot DA white dwarfs (cf.\ Fig.~\ref{f:HR}). However, the
existence of PNe around these stars is very difficult to understand. The
theoretical post-AGB ages of 47000\,K and 38000\,K white dwarfs amount to
$\approx 3\cdot 10^{6}$\,years, and $\approx 6\cdot 10^6$ years, respectively.
That is much older than any PNe found in our sample (cf.\ Table~\ref{t:cpn})
and one would not expect a PN to survive that long. Moreover, the nebulae of
\object{HDW\,4} and \object{HaWe\,5} do not appear to be extraordinarily old.
Their kinematical ages amount to only 2000 and 3000 years and their surface
brightnesses are not atypically low. Actually, the nebula of \object{HaWe\,5}
is bright enough to contaminate the cores of the \halpha\ and \hbeta\ lines in
the central star spectrum (Fig.~\ref{f:dagitter}). 

We are also not aware of any binary evolution models which can explain the
properties of these two central stars and their nebulae, because no scenario is
able to produce a remnant which reaches the positions of \object{HDW\,4} and
\object{HaWe\,5} within a reasonable time after departure either from the AGB
or RGB (see e.g.\ Iben \& Tutukov \cite{IT85}; Iben \cite{I86}; Iben \& Tutukov
\cite{IT93}). The fundamental reason is that the time scales in the white dwarf
domain are determined by the gravitational-internal energy of the stellar
remnant (Bl\"ocker \& Sch\"onberner \cite{BS90}). If the stellar evolution is
abbreviated by a companion stripping off matter from the primary, the structure
of the resulting remnant is less compact than that of a normal post-AGB star of
the same mass. That means the gravitational-internal energy content is larger
and cooling needs even more time. 

Since it seems impossible to explain the existence of a PN around these stars,
we investigated the hypothesis that the nebula \object{HDW\,4} and
\object{HaWe\,5} only mimic the appearance of a PN. It is well known that
white dwarfs in cataclysmic binaries can eject shells during an outburst caused
by a thermonuclear runaway after they have accreted a certain amount of
hydrogen-rich material from the companion, usually a late type main sequence
star. The mass of the ejected material is of the order $10^{-4}M_\odot$ (cf.\
Table 5.7 of Warner \cite{W95}), much smaller than the mass of a typical PNe.
Sometime after this nova event the expanding shell can become visible as
spatially resolved object with a morphology not much different from planetary
nebulae (see e.g.\ Slavin et al. \cite{SOD95}; Tweedy \cite{T95}). We inverted
the Shklovsky formula (see e.g.\ Eq.~1 in Paper~III) and estimated shell masses
of \object{HDW\,4} and \object{HaWe\,5} of $10^{-3}M_\odot$ and $2\cdot
10^{-4}M_\odot$, respectively, consistent with a nova interpretation. 

However, nova shells usually have expansion velocities of 500\,km/s or higher
(see Slavin et al.\  \cite{SOD95} and references therein). We have obtained a
spectrum of the \halpha\ region of \object{HDW\,4} with a spectral resolution
of $1.63\pm 0.02$\,\AA\ (FWHM; measured from Gaussian fits of comparison lamp
lines). We applied the same fit procedure to the nebular \halpha\ line and the
\ion{N}{ii} 6548/84\,\AA\ lines and got an average width of $1.85\pm
0.06$\,\AA. If we ignore thermal broadening this provides us with an upper
limit on the expansion velocity of $2v_{\mathrm{exp}} < 47$\,km/s. That
corresponds to typical PN expansion velocities, and is much lower than the
usual velocities observed in nova shells, even if one considers projection
effects. 

Until now we did not detect any binary companion. The low resolution spectra of
\object{HDW\,4} and \object{HaWe\,5} end at 6800\,\AA\ and don't display any
signs of a cool companion. Ongoing mass accretion from a companion must be very
low if any, because otherwise the atmospheres of \object{HDW\,4} and
\object{HaWe\,5} would be contaminated by helium and heavy elements and would
not show the pure hydrogen spectrum we observe. A probable counterpart of our
mysterious stars might be the very old nova \object{CK\,Vul} which exploded in
1670. Shara et al.\ (\cite{SMW85}) recovered it as a very faint star inside a
faint nebulosity on deep CCD images. From the size of the nebulosity Shara et
al.\ estimated an expansion velocity of 59\,km/s, lower than for any other
known nova shell. However, a direct measurement of line widths is still
lacking. Naylor et al.\ (\cite{NCM92}) took the nova nature of \object{CK\,Vul}
into question because the proposed central star doesn't show \halpha\ in
emission as it is usually observed even in old nova systems (Ringwald et al.\
\cite{RNM96}). Shara et al.\ (\cite{SMW85}) concluded that any mass transfer 
in the \object{CK\,Vul} system must virtually have stopped. An infrared
investigation gave evidence for cool dust in the system (Harrison \cite{H96}).
Although observational evidence is spurious it may well be that
\object{HDW\,4}, \object{HaWe\,5}, and \object{CK\,Vul} belong to the same
class of objects. Theoretical calculations (Shara et al.\ \cite{SPK93};
Prialnik \& Kovetz \cite{PK95}) predict that nova outbursts of white dwarfs
with ``atypical low masses'' of 0.60 or $0.65M_\odot$ (roughly the masses of
\object{HDW\,4} and \object{HaWe\,5}) should be very slow and eject high shell
masses compared to nova with more ``typical'' white dwarf masses
($0.91M_\odot$; Webbink \cite{W90}). This might explain the observational
properties of \object{HDW\,4} and \object{HaWe\,5} discussed above. However, no
definite conclusions are possible without further observational data. A deep IR
search for companions in both CSPN will be of special importance. 

\subsection{Central stars of hybrid spectral type}

\begin{figure}
\epsfxsize=5.4cm
\epsffile{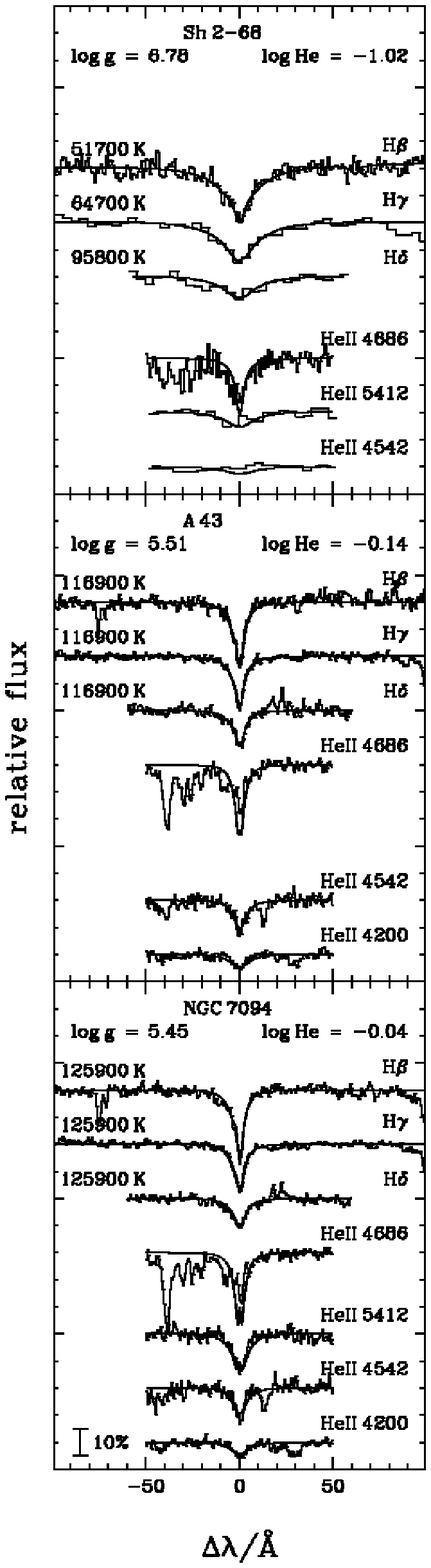}
\caption{Fits of the hybrid PG\,1159 stars. The \ion{He}{ii} 4686\,\AA\
lines of \object{A\,43} and \object{NGC\,7094} were not used for the fit. The observed
and theoretical spectra are only shown for comparison purposes}
\label{f:pg1159gitter}
\end{figure}

Our sample contains three hydrogen-rich PG\,1159 stars or hybrid CSPN
(\object{Sh\,2-68}, \object{A\,43}, \object{NGC\,7094}). Their spectra show a
\ion{C}{iv} absorption trough in the region of the \ion{He}{ii} 4686\,\AA\ line
as it is typical for the hydrogen-poor, carbon- and helium-rich PG\,1159 stars
(see Dreizler et al.\ \cite{DWH95} for a review). However, unlike in PG\,1159
stars the Balmer lines are visible, too, which indicates a much higher amount
of hydrogen in their atmospheres. The first star of this type,
\object{Sh\,2-68}, was found during our survey of central stars of old PNe
(Paper~II). Dreizler et al.\ (\cite{DWH95}) analyzed \object{A\,43} and
\object{NGC\,7094} taking also the lines of C, N, and O into account. They
derived $\teff = 110000$\,K and $\log g = 5.7$ for both stars. The agreement
with our analysis of \object{A\,43} is good, while we derive a significantly
higher temperature for \object{NGC\,7094}. The difference of the Balmer lines
of both stars (Fig.~\ref{f:pg1159gitter}) indicates that the parameters of
these stars cannot be equal. \object{NGC\,7094} is likely to be hotter and/or
of lower gravity than \object{A\,43}. \object{HS\,2324+3944} is the only star
of this class without a PN (Dreizler et al.\ \cite{DWH96}). A search for a
nebula around this star was not successful (Werner et al.\ \cite{WBR97}). 

The hybrid CSPN (hydrogen-rich PG\,1159 stars) \object{A\,43} and
\object{NGC\,7094} are placed on the horizontal part of the post-AGB tracks,
where only one other CSPN of our sample is found. All other stars are more
evolved. The comparison with the evolutionary tracks yields post-AGB ages
smaller than 1000\,yrs, which is short compared to the kinematical ages of
their PNe (25000\,yrs and 11000\,yrs for \object{A\,43} and \object{NGC\,7094},
respectively). However, if we consider the error bars and uncertainties of age
determinations both ages can in principle brought into agreement. On the other
hand their position in the HR diagram, distinct from the other central stars of
old PNe, argues for the reality of this age discrepancy. The absence of a PN
around \object{HS\,2324+3944} ($\teff = 130000$\,K; $\log g = 6.2$; Dreizler et
al.\ \cite{DWH96}) is another argument for the reality of the discrepancy.
\object{A\,43} and \object{NGC\,7094} are good candidates for born again
central stars (Sch\"onberner \cite{S79}, \cite{S83}; Iben et al.\
\cite{IKT83}), which returned to the AGB after a late thermal flash and repeat
the post-AGB evolution, now as helium-burning stars. The result is a seemingly
young central star in an old PNe. This explains the age discrepancy we found
for the three hybrid stars discussed above. Fittingly, this scenario is often
invoked to explain the chemical abundances of PG\,1159 stars. However, we found
another central star (\object{DeHt\,2}) with very similar temperature and
gravity and an old PNe ($t_{\mathrm{kin}} = 27000$\,yrs). If the age
discrepancy found for the hybrid central stars is explained by the born-again
scenario it should apply to \object{DeHt\,2} as well, but no sign of a chemical
enrichment of the atmosphere is present. This star might indicate that born
again evolution is not necessarily linked to strong chemical processing of the
photosphere.

\begin{figure}
\epsfxsize=5.4cm
\epsffile{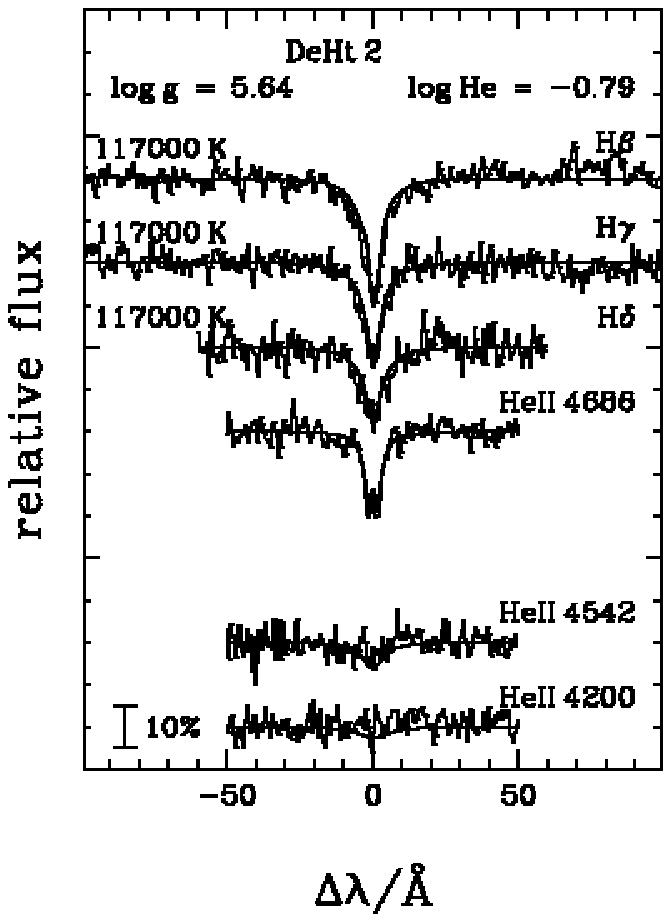}
\caption{Fit of the high luminosity central star \object{DeHt\,2}}
\label{f:dhw2gitter}
\end{figure}

\subsection{Comments on individual objects}

\paragraph{\object{PN\,G 025.4$-$04.7} (\object{IC\,1295}):} The PN is
relatively bright and produces severe contamination of the core of \hbeta\ and
the \ion{He}{ii} 4686\,\AA\ line. Thus we excluded 4686\,\AA\ from the fit and
estimated the helium abundance from the \ion{He}{ii} lines at 5412\,\AA\ and
4542\,\AA. Only the wings of the \hbeta\ line were fitted. 
 
\paragraph{\object{PN\,G 036.0+17.6} (\object{A\,43}):} The \ion{He}{ii}
4686\,\AA\ region is heavily contaminated by \ion{C}{iv} lines and was excluded
from the fit. 

\paragraph{\object{PN\,G 036.1$-$57.1} (\object{NGC\,7293}):} The central star 
was previously analysed by M\'endez et al.\ (\cite{MKH88a}). Their result was
$\teff = 89000$\,K and $\log g = 6.9$. The gravity is virtually identical with
our value $\log g = 7.00 \pm 0.22$, but their temperature is somewhat lower
($\teff = 103600 \pm 5500$\,K). This is easily explained when one takes into
account that \hgamma\ was the only analysed hydrogen line. Our fit of \hgamma\
results in $\teff = 89000$\,K, in perfect agreement with the M\'endez et al.\
result. 

\paragraph{\object{PN\,G 047.0+42.4} (\object{A\,39}):} The McCarthy et al.\
(\cite{MMK97}) result $\teff = 110000\,K$, $\log g = 6.3$, and $\nhe/\nh =
0.11$ agrees with our parameters. 

\paragraph{\object{PN\,G 60.8$-$03.6} (\object{NGC\,6720}):} The cores of
\hbeta, \hgamma, and \ion{He}{II} 4686\,\AA\ are contaminated by the nebula 
lines and were excluded from the fit as well as the interstellar \ion{Ca}{II} H
and K absorption in \hepsilon.  McCarthy et al.\ (\cite{MMK97})  fitted a high 
resolution spectrum obtained with the HIRES echelle spectrograph of  the
Keck\,I telescope and derived a somewhat low temperature of the  central star
(80000\,K). High resolution spectra are certainly useful  if one has to deal
with bright nebulae. On the other hand it is  difficult to reduce the very
broad Balmer lines of a $\log g = 7.0$  star, which are distributed over
several orders of the HIRES echelle  spectrum, without deformation of the line
profiles.

\paragraph{\object{PN\,G 066.7$-$28.2} (\object{NGC\,7094}):} The \ion{He}{ii} 
4686\,\AA\ region is heavily contaminated by \ion{C}{iv} lines and was excluded
from the fit. 

\paragraph{\object{PN\,G 125.9$-$47.0} (\object{PHL\,932}):} The temperatures
derived from the Balmer lines and the ionization equilibrium of helium differ
by 3000\,K. This is certainly caused by our neglect of metal line blanketing.
The Balmer line profile differences of the model spectra and the observed
spectra are in close agreement to the expectation from the theoretical
comparison in Fig.~10 of Napiwotzki (\cite{N97}). M\'endez et al.\
(\cite{MGH88b}) derived $\teff = 37000\pm 2000$\,K, $\log g = 5.5\pm 0.2$ which
is in marginal agreement with our results. Part of the difference might be
explained by the use of different model atmospheres and fitting procedures. 

\paragraph{\object{PN\,G 158.9+17.8} (\object{PuWe\,1}):} The McCarthy et al.\
(\cite{MMK97}) derive $\teff = 110000\,K$, $\log g = 7.0$, and $\nhe/\nh =
0.02$. The temperature is slightly higher than our result (94000\,K). See
discussion of \object{NGC\,6720}. 

\paragraph{\object{PN\,G 197.4$-$06.4} (\object{WeDe\,1}):} \object{WeDe\,1}
was first announced as a very hot, high-gravity star by Napiwotzki \&
Sch\"onberner (\cite{NS91b}). A LTE analysis of the moderate quality spectrum
available at that time resulted in lower limits on $\teff$ ranging from
100000\,K to 200000\,K (depending on the adopted helium abundance) and a lower
gravity limit of $\log g = 8.5$. A NLTE analysis using a better spectrum and
model atmospheres similar to that described in the present paper yielded in
$\teff = 217000$\,K and $\log g = 7.35$ (Napiwotzki \cite{N95}). Independently,
Liebert et al.\ (\cite{LBT94}) performed a LTE analysis and derived $\teff =
99000\,\mathrm{K} \ldots 165000$\,K, again depending on the adopted helium
abundance. Since the quality of the Liebert et al.\ spectrum is superior to
that of our Calar Alto spectrum, we used their spectrum for our analysis. The
result is $\teff = 141000\pm 32000$\,K and $\log g = 7.53 \pm 0.32$. The
dependence of the fit result on the adopted helium abundance is much lower for
our NLTE analysis than it is for LTE analyses (cf.\ Napiwotzki \cite{N97}).

\paragraph{\object{PN\,G 215.5$-$30.8} (\object{A\,7}):} The McCarthy et al.\
(\cite{MMK97}) result $\teff = 100000\,K$, $\log g = 6.6$, and $\nhe/\nh =
0.05$ agrees with our parameters. 

\section{Helium abundances and the onset of diffusion}
\label{s:helium}

A wide spread of helium abundances is observed in the atmospheres of
hydrogen-rich central stars. Vennes et al.\ (\cite{VPF88}) have shown that
helium sinks down due to gravitational settling in the atmospheres of
hydrogen-rich white dwarfs. Diffusive equilibrium is reached fast and no
visible traces of helium should be present in these atmospheres. Therefore it
was considered the most consistent physical picture of DAO white dwarfs that
their atmospheres should consist of an extremely thin hydrogen layer on top of
the helium envelope. The helium lines should form in the deeper, helium-rich
regions. In the context of the Fontaine \& Wesemael (\cite{FW87}) scenario of
the spectral evolution of white dwarfs DAO stars are transition objects from
the helium-rich DO to the hydrogen-rich DA sequence, which have just built up
an extremely thin hydrogen-layer. The very thin layer hypothesis can be tested
by detailed fits of the \ion{He}{ii} 4686\,\AA\ profiles. The formation depth
of this line is quite different for stratified and homogeneous atmospheres: the
line is much broader and shallower in the stratified case. Napiwotzki \&
Sch\"onberner (\cite{NS93}) demonstrated that the \ion{He}{ii} 4686\,\AA\ line
of \object{Sh\,2-216} is well fitted by a homogeneous model atmosphere while a
stratified atmosphere could be excluded. Figs.~\ref{f:daogitter1}
and~\ref{f:daogitter2} show also good agreement between the profiles computed
from our homogeneous models and those observed for \ion{He}{ii} 4686\,\AA.
Bergeron et al.\ (\cite{BWB94}) extended this investigation to their DAO sample
and found also that the atmospheres of most stars are not stratified. However,
they presented one example, \object{PG\,1305$-$017}, with a probably stratified
atmosphere, which might, indeed, be an object caught during the transition from
DO to DA spectral type. 

\begin{figure}
\epsfxsize=8.5cm
\epsffile[0 18 750 580]{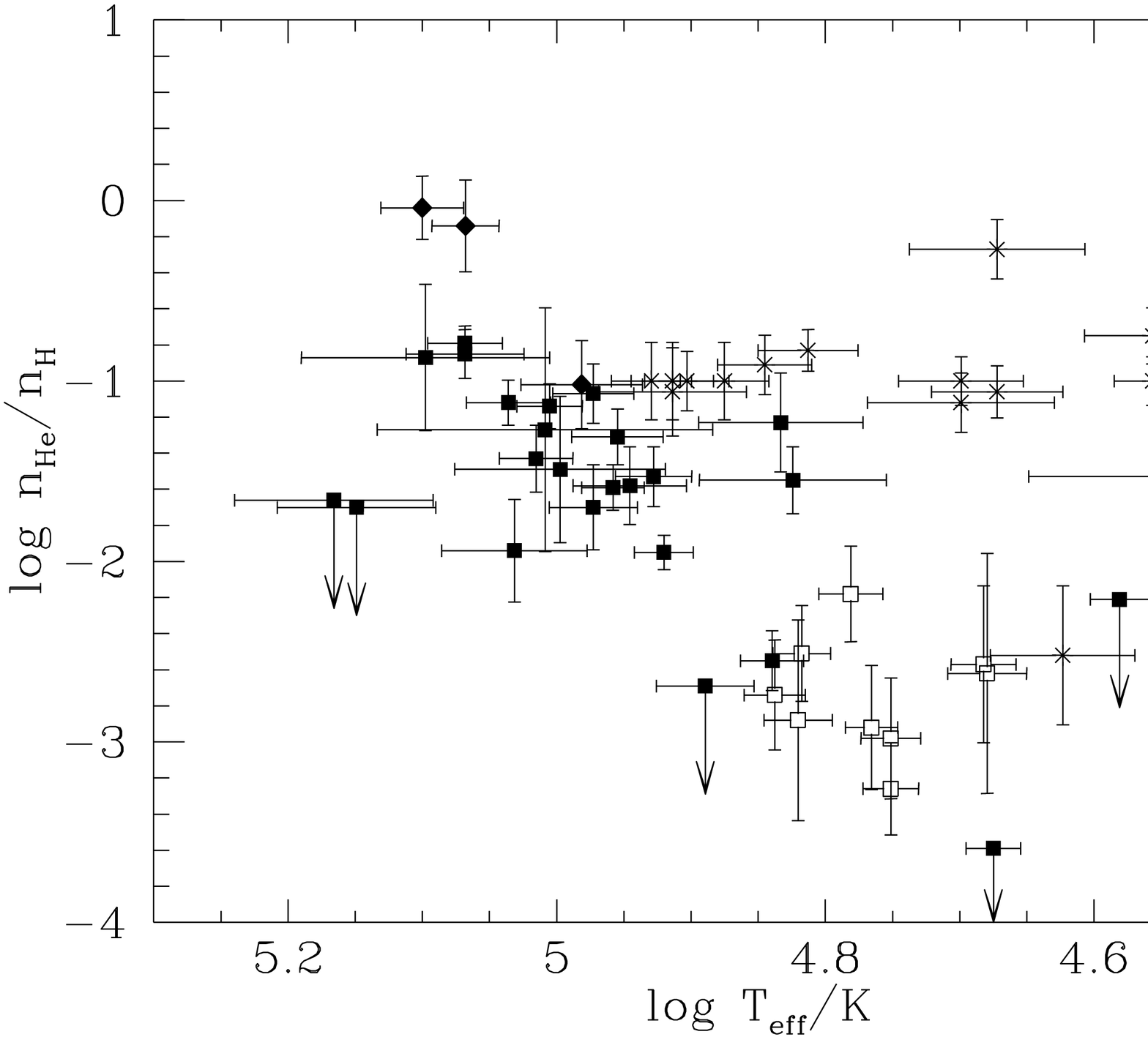}

\epsfxsize=8.5cm
\epsffile[0 18 750 580]{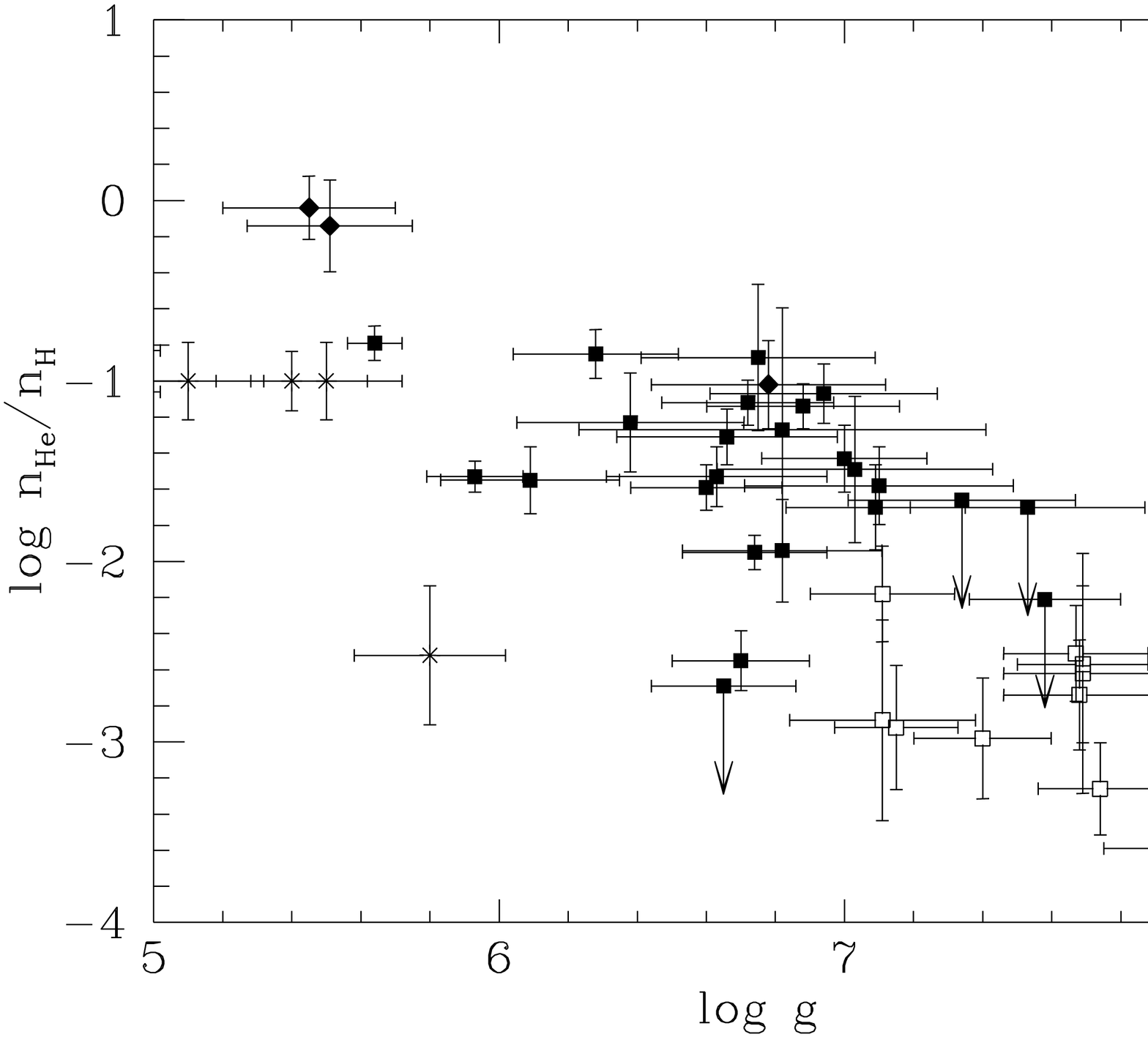}

\epsfxsize=8.5cm
\epsffile[0 18 750 580]{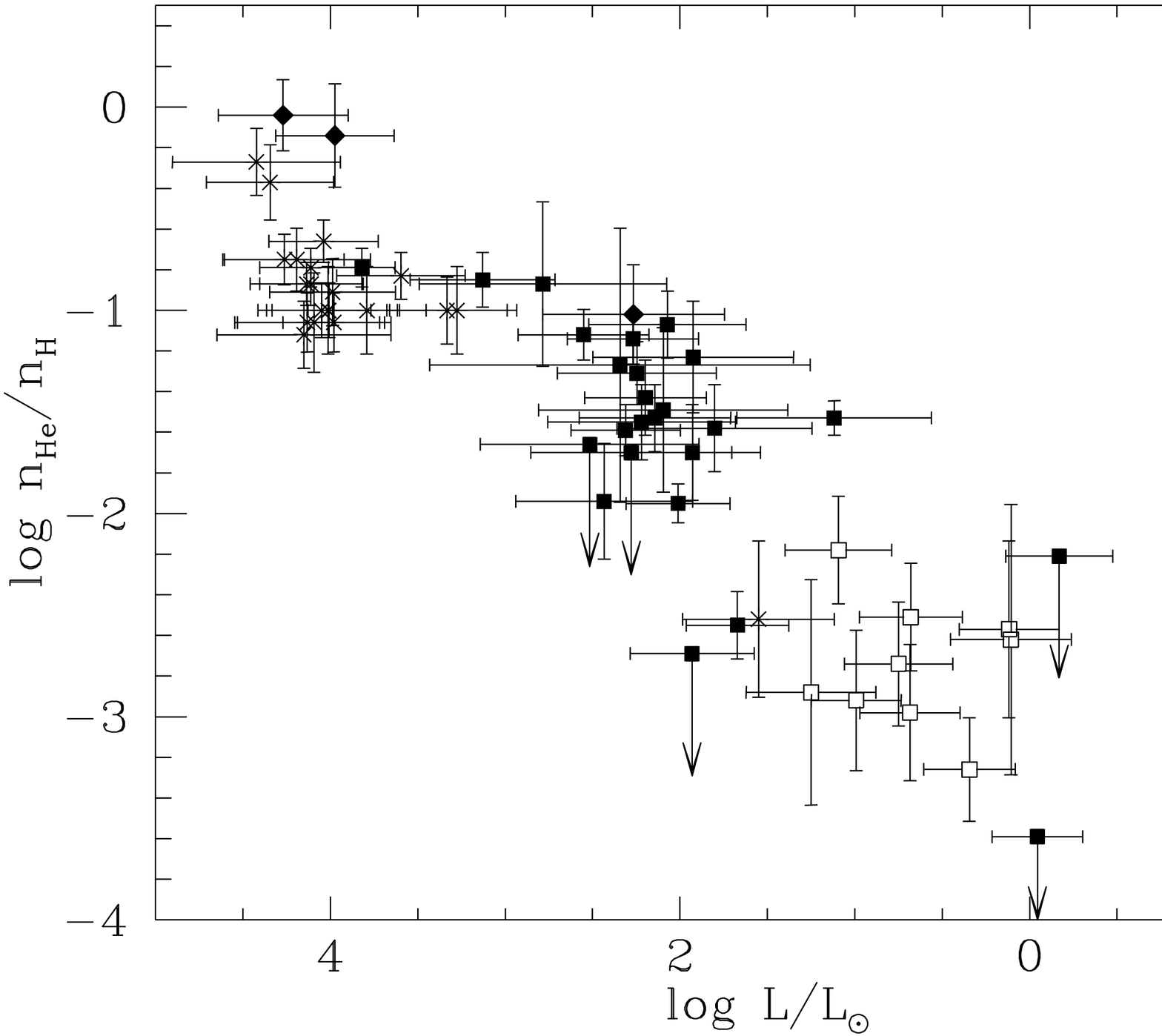}
\caption[]{Photospheric helium abundance as function of effective temperature,
gravity, and luminosity (top to bottom). Our results (filled squares; hybrid
CSPN: filled diamonds) are supplemented with the DAO white dwarfs analysed by
Bergeron et al.\ (\cite{BWB94}; open squares) and the CSPN analysis of M\'endez
et al. (\cite{MKH88a}; stars). Upper limits are marked by arrows} 
\label{f:helium}
\end{figure}

Napiwotzki (\cite{N93b}, \cite{N95}) and Bergeron et al.\ (\cite{BWB94}) noted
a correlation between helium abundance and gravity. However, Bergeron et al.\
(\cite{BWB94}) did not consider it to be significant and stressed a stronger
correlation with $\teff$. We search for correlations between the helium
abundance and temperature, surface gravity and luminosity in
Fig.~\ref{f:helium}. We combine our results with the DAO analyses of Bergeron
et al.\ (\cite{BWB94}) and the CSPN analysis of M\'endez et al.\
(\cite{MKH88a}; \cite{MKH92}).  All stars of the M\'endez et al.\ sample (with
the exception of the peculiar objects \object{EGB\,5} and \object{PHL\,932})
are on the constant luminosity part of the central star tracks. Some stars
exhibit helium enrichment by dredge up of nuclear processed material, but none
shows a significant depletion below the solar value. If we ignore these
luminous M\'endez et al.\ stars, indeed a correlation of helium abundance with
temperature becomes visible (upper panel of Fig.~\ref{f:helium}), however, not
a very good one. On the one hand the hottest stars \object{WeDe\,1} and
\object{EGB\,1} have upper limits below the solar value and on the other hand
several of the cooler stars are only mildly depleted. A better correlations is
found between $\nhe/\nh$ and the surface gravity (middle panel). However, in
this case there are exceptions from the general trend present, too. A far
better correlation is found between helium abundance and luminosity (lower
panel). All stars, whether of post-AGB nature or not, can be explained by one
trend line, although the scatter seems to be larger than expected from the
observational errors alone. For all stars with a luminosity higher than
$\approx$$300L_\odot$ the helium abundance is close to the ``solar'' value
$\nhe/\nh = 0.1$ or above. For stars with lower luminosity the helium abundance
steadily decreases with decreasing luminosity. Since effective temperature and
surface gravity are related to luminosity it is not surprising that a
correlation with temperature is found in a sample with roughly equal gravity
(Bergeron et al.\ \cite{BWB94}), or a correlation with gravity in a sample with
roughly equal temperature (Napiwotzki \cite{N93b}; \cite{N95}). Although
intrinsic scatter seems to be present in the correlation of helium abundance
with luminosity its existence is remarkable. The observed abundances of heavy
elements in hot white dwarfs usually display a more or less stochastical
dependence on stellar parameters without any clear trend (cf.\ Chayer et al.\
\cite{CFW95}). 

Recent theoretical calculations performed by Unglaub \& Bues (\cite{UB98}) show
that the helium abundances observed in DAO stars can be explained if a
reasonable mass loss rate is adopted. Mass loss counteracts the gravitational
settling of helium and slows it down, if the strength of the stellar wind is
high enough. The mass loss of hot stars is radiatively driven and a strong
function of luminosity (e.g.\ Abbott \cite{A82}). Thus a correlation between
luminosity and helium abundance is also expected on theoretical grounds. The
calculations of Unglaub \& Bues (\cite{UB98}) show that the static equilibrium
abundances are not reached within typical evolutionary time scales of hot white
dwarfs. Therefore the stellar abundances are the result of a dynamical
equilibrium between diffusion time scales and the evolution of the star. This
might explain why sometimes stars with similar parameters show different
photospheric abundances. An example is given by the DA \object{DeHt\,5} and the
DAO \object{Sh\,2-174}. \object{DeHt\,5} has a slightly higher temperature and
slightly lower gravity than \object{Sh\,2-174}. Thus we would expect the helium
abundance of \object{DeHt\,5} to be higher than in \object{Sh\,2-174}, but the
spectrum \object{Sh\,2-174} shows an easily detectable \ion{He}{ii} 4686\,\AA\
line (Fig.~\ref{f:daogitter1}) while the line is below the detection limit in
the spectrum of \object{DeHt\,5} (Fig.~\ref{f:dagitter}). However, if the
helium abundance is the result of time dependent diffusion it should depend on
the previous evolution of the stars. Thus stars with similar parameters but
different evolutionary histories can show different abundance patterns. We
conclude that today observational evidence and theoretical calculations of the
evolution of surface helium abundance show semi-quantitative agreement. What is
needed for an improved understanding are self-consistent diffusion calculations
taking the stellar evolution into account and an extension of the comparison to
as many heavy elements as possible. 

\section{Conclusions}

We presented a NLTE model atmosphere analysis of 27 hydrogen-rich central stars
of old PNe. The stars were taken from Paper~III in this series, where we
provided classifications for a total of 38 CSPN. The analysis of the H-rich
CSPN is hampered by the Balmer line problem: for most stars a consistent fit to
all Balmer lines was not possible. Generally the highest Balmer line yields the
highest temperature. The likely solution of this problem was presented by
Werner (\cite{W96}). He demonstrated that the inclusion of Stark broadening for
C, N, and O lines can have a strong influence on the atmospheric structure of
very hot hydrogen-rich stars. The lower Balmer lines are most strongly affected
and their analysis with standard NLTE model grids yields unreliable
temperatures. It is likely that the effect on the emergent spectrum is
pronounced enough to solve the Balmer line problem. Since the computation of a
NLTE model grid accounting for the influence of C, N, and O in the above
described manner requires enormous amounts of computer time, we analyzed the
CSPN with our atmospheres containing only H and He and derived $\teff$ from the
fit of \hdelta\ (or \hepsilon\ if available). 

A previously reported gap in the hydrogen-rich evolutionary sequence is filled
by our sample and, therefore, shown to result from a selection effect. Taking
into account the hot white dwarfs analyzed by Bergeron et al.\ (\cite{BWB94})
we have now a continuous sequence from the CSPN to the white dwarfs. The
observational findings are well explained by a two channel scenario with a
hydrogen-rich and a hydrogen-poor sequence. It is no longer necessary to invoke
an one channel scenario (Fontaine \& Wesemael \cite{FW87}) to explain the
spectral evolution of pre-white dwarfs. Note that our findings do not rule out
that {\em some} white dwarfs change their spectral type from DO to DA (see
discussions in Fontaine \& Wesemael \cite{FW97} and Napiwotzki \cite{N98}). 

Stellar masses were computed from a comparison with evolutionary tracks and a
mass distribution for the hydrogen-rich central stars of old PNe was derived.
The distribution has a sharp peak centered at $0.55M_\odot$ This value and the
general shape are similar to the results of previous analysis of hot DA white
dwarfs (Bergeron et al.\ \cite{BSL92}; Finley et al.\ \cite{FKB97}; Napiwotzki
et al.\ \cite{NGS99}) and of a recent distance independent investigation of PNe
(Stasi\'nska et al. \cite{SGT97}). A relatively large fraction of low mass
central stars is present in our sample. This might be explained by our
selection criteria which tend to prefer the old and thin nebulae around slowly
evolving post-RGB stars. 

Although the majority of analyzed stars is well explained by standard post-AGB
evolution, there exists a number of stars for which other scenarios have to be
invoked. The properties of three stars are probably best explained by born
again evolution. Two of these are hybrid CSPN (hydrogen-rich PG\,1159 stars),
but surprisingly the third star doesn't show any signs of chemical enrichment
in its atmosphere. The third hybrid CSPN in our sample (\object{Sh\,2-68}) has
parameters which place it among the group of normal DAO central stars. 

The fundamental parameters of five stars are not in accordance with a standard
post-AGB evolution. They are likely the outcome of a close binary evolution.
Several scenarios are possible. Four stars are reasonably well explained by the
evolution of stars whose hydrogen-envelope has been stripped off by a nearby
companion during the first red giant branch phase. The remnants will eventually
become low mass white dwarfs with a degenerate helium core. Although the
post-RGB evolution is considerably slower than post-AGB evolution the existence
of low density nebulae around these objects is in accordance with theoretical
time scales. The kinematical ages of the observed PNe are consistent with the
evolutionary post-RGB age, except for \object{PHL\,932}, where this scenario is
ruled out by a very large discrepancy between predicted post-RGB age and
kinematical age of the nebula. Two stars (\object{HDW\,4} and \object{HaWe\,5})
remain mysterious. They resemble ordinary hot DA white dwarfs with evolutionary
ages of several million years. However, this large age is in clear
contradiction with the presence of a nebula. Close binary evolutionary
scenarios do not resolve this discrepancy, because the remnants need several
million years after the possible ejection of a nebula to reach the position of
\object{HDW\,4} and \object{HaWe\,5} in the HR-diagram. We speculate that the
nebulae of these stars are produced by nova-like events. However, no direct
observational evidence can be given in the moment.

Napiwotzki \& Sch\"onberner (\cite{NS93}) and Bergeron et al.\ (\cite{BWB94})
have shown that the atmospheres of DAO stars are chemically homogeneous, i.e.\
not layered. We have shown that a good correlation between helium abundance and
luminosity is present. Depletion starts when the stellar luminosity falls below
$L\approx 300 L_\odot$,  and the helium abundance steadily decreases with
decreasing luminosity. Although the scatter is probably larger than expected
from the observational errors alone, the correlation is surprisingly tight
considering the more or less stochastic abundance variations usually found in
hot white dwarfs (Chayer et al.\ \cite{CFW95}). The existence of the
correlation is in qualitative agreement with recent theoretical calculations of
gravitational settling in the presence of a stellar wind (Unglaub \& Bues
\cite{UB98}).

\acknowledgements
The author thanks T.~Rauch, U.~Heber, and S.~Moehler who made the
medium-resolution observations, and H.~Edelmann, who did the data reduction for
one run. I gratefully acknowledge the help of U.~Kolb, H.~Drechsel, and
K.~Schenker, who answered many silly questions about cataclysmic binaries. 
This work was supported by DFG travel grants and by DARA grant 
50\,OR\,96029-ZA.

\appendix
\section{A compilation of CSPN and white dwarf parameters from literature}

Parameter determinations of CSPN and hot white dwarfs and related objects are
widely scattered through literature. Many sources are not easily accessible.
For this reason we provide a table with parameters and references of the
objects displayed in Fig.~\ref{f:HR}. This compilation may be useful for future
investigations. The most obvious application is the selection of candidates for
deep imaging to look for faint PNe. We tried to collect all data for central
stars and pre-white dwarfs with temperatures higher than 30000\,K from
literature, which were analyzed with state-of-the-art techniques. Stars already
analyzed in this paper (cf.\ Table~\ref{t:cpn}) are omitted from
Table~\ref{t:lit}. We used the same lower temperature limit for the white
dwarfs, but did not aim for completeness for the cooler DA white dwarfs ($\teff
< 60000$\,K). The results are provided in Table~\ref{t:lit}. The stars are
grouped according to their spectral classes and their temperature. References
are given in Table~\ref{t:ref}. Objects in globular clusters were excluded from
this compilation. 

While the DAO and DA analyses are usually based on LTE model atmospheres (with
the exception of the Napiwotzki et al.\ \cite{NGS99} analysis), NLTE analyses
were performed for all other stars listed in Table~\ref{t:lit}. Since
deviations from LTE produce significant effects in the hottest DA and DAO white
dwarfs (Napiwotzki \cite{N97}; Napiwotzki et al.\ \cite{NGS99}), one should be
aware of possible systematic effects. 

\newcommand{\BOK}{1}
\newcommand{\BH}{2}
\newcommand{\BWB}{3}
\newcommand{\DrH}{4}
\newcommand{\DW}{5}
\newcommand{\DWH}{6}
\newcommand{\DHN}{7}
\newcommand{\DWHR}{8}
\newcommand{\FKB}{9}
\newcommand{\HWD}{10}
\newcommand{\HMM}{11}
\newcommand{\HDW}{12}
\newcommand{\HKH}{13}
\newcommand{\Hus}{14}
\newcommand{\HBH}{15}
\newcommand{\LB}{16}
\newcommand{\LTN}{17}
\newcommand{\M}{18}
\newcommand{\MMK}{19}
\newcommand{\MKHHG}{20}
\newcommand{\MKH}{21}
\newcommand{\Nap}{22}
\newcommand{\NHJ}{23}
\newcommand{\NGS}{24}
\newcommand{\PRB}{25}
\newcommand{\R}{26}
\newcommand{\RHH}{27}
\newcommand{\RDW}{28}
\newcommand{\RW}{29}
\newcommand{\RKN}{30}
\newcommand{\Rey}{31}
\newcommand{\SWW}{32}
\newcommand{\TK}{33}
\newcommand{\WWP}{34}
\newcommand{\WBR}{35}
\newcommand{\ZW}{36}

\begin{table}
\caption{References for Table~\ref{t:lit}}
\label{t:ref}
\begin{tabular}{ll}
\BOK	&Barstow et al. (\cite{BOK95})\\
\BH	&Bauer \& Husfeld (\cite{BH95})\\
\BWB	&Bergeron et al.\ (\cite{BWB94})\\
\DrH	&Dreizler \& Heber (\cite{DH98})\\
\DW	&Dreizler \& Werner (\cite{DW96})\\
\DWH	&Dreizler et al.\ (\cite{DWH95})\\
\DHN	&Dreizler et al.\ (\cite{DHN95})\\
\DWHR	&Dreizler et al.\ (\cite{DWH99})\\
\FKB	&Finley et al.\ (\cite{FKB97})\\
\HWD	&Heber et al.\ (\cite{HWD88})\\
\HMM	&Herrero et al.\ (\cite{HMM90})\\
\HDW	&Hoare et al.\ (\cite{HDW96})\\
\HKH	&Homeier et al.\ (\cite{HKH98})\\
\Hus	&Husfeld D. (\cite{Hus87})\\
\HBH	&Husfeld et al.\ (\cite{HBH89})\\
\LB	&Liebert \& Bergeron (\cite{LB95})\\
\LTN	&Liebert et al.\ (\cite{LTN95})\\
\M	&McCarthy (\cite{M88})\\
\MMK	&McCarthy et al.\ (\cite{MMK97})\\
\MKHHG	&M\'endez et al.\ (\cite{MKH88a})\\
\MKH	&M\'endez et al.\ (\cite{MKH92})\\
\Nap	&Napiwotzki (\cite{N93a})\\
\NHJ	&Napiwotzki et al.\ (\cite{NHJ95})\\
\NGS	&Napiwotzki et al.\ (\cite{NGS99})\\
\PRB	&Pe\~na et al.\ (\cite{PRB97})\\
\R	&Rauch (\cite{R93})\\
\RHH	&Rauch et al.\ (\cite{RHH91})\\
\RDW	&Rauch et al.\ (\cite{RDW98})\\
\RW	&Rauch \& Werner (\cite{RW99})\\
\RKN    &Rauch et al. (\cite{RKN99}\\
\Rey	&Reynolds (\cite{R87})\\
\SWW	&Saurer et al.\ (\cite{SWW97})\\
\TK	&Tweedy \& Kwitter (\cite{TK94})\\
\WWP	&Werner et al.\ (\cite{WWP96})\\
\WBR	&Werner et al.\ (\cite{WBR97})\\
\ZW	&Zanin \& Weinberger (\cite{ZW97})\\
\end{tabular}
\end{table}

\begin{table*}
\begin{minipage}{17cm}
\caption{Fundamental parameters of hot CSPN, white dwarfs, and related
objects compiled from the literature}
\label{t:lit}
\scriptsize
\begin{tabular}{l@{ }rl@{ }c@{ }l|}
Name	&$\teff$	&$\log g$ &PN	&Ref.\\ \hline
\multicolumn{5}{l|}{O(H) stars}\\
\object{K\,1-22}         &141.0  &6.73   &+      &\RKN\\
BlDz\,1         &128.0  &6.85   &+      &\RKN\\
\object{A\,20}           &119.0  &6.13   &+      &\RKN\\
\object{NGC\,2438}       &114.0  &6.62   &+      &\RKN\\
\object{NGC\,1360}	&110.0	&5.60	&+	&\HDW\\
\object{A\,15}		&110.0	&5.70	&+	&\MMK\\
\object{BE UMa}		&105.0	&6.50	&+	&\LTN\\
\object{LSS\,1362}	&100.0	&5.30	&+	&\HWD\\
\object{NGC\,2022}	&100.0	&5.30	&+	&\MMK\\
\object{MeWe\,1-3}	&100.0	&5.50	&+	&\SWW\\
\object{IC\,289}		&100.0	&5.60	&+	&\MMK\\
\object{NGC\,2610}	&100.0	&5.80	&+	&\MMK\\
\object{HaTr\,7}		&100.0	&6.00	&+	&\SWW\\
\object{EGB\,6}          &100.0  &7.00   &+      &\LB\\
\object{A\,36}		&95.0	&5.30	&+	&\M\\
\object{DHW\,1-2}	&90.0	&5.00	&+	&\SWW\\
\object{Lo\,8}		&90.0	&5.10	&+	&\HMM\\
\object{LSS\,2018}	&90.0	&5.10	&+	&\HMM\\
\object{J\,320}		&85.0	&4.70	&+	&\MMK\\
\object{LSE\,125}	&85.0	&5.10	&+	&\MKHHG\\
\object{NGC\,7009}	&82.0	&4.90	&+	&\MKH\\
\object{NGC\,4361}	&82.0	&5.50	&+	&\MKH\\
\object{BD+28$^\circ$4211}&82.0	&6.20	&?	&\Nap,\ZW\\
\object{NGC\,3242}	&75.0	&4.75	&+	&\MKH\\
\object{KS\,292}		&75.0	&5.00	&$-$	&\RHH\\
\object{Feige\,67}	&75.0	&5.20	&$-$	&\BH\\
\object{NGC\,1535}	&70.0	&4.65	&+	&\MKH\\
\object{IC\,2448}	&65.0	&4.80	&+	&\MKH\\
\object{NGC\,6058}	&65.0	&4.80	&+	&\HMM\\
\object{Vy\,1-1}		&60.0	&4.20	&+	&\MMK\\
\object{NGC\,6210}	&50.0	&3.90	&+	&\M\\
\object{NGC\,6891}	&50.0	&4.00	&+	&\MKH\\
\object{NGC\,6826}	&50.0	&4.00	&+	&\MKH\\
\object{IC\,4637}	&50.0	&4.05	&+	&\MKH\\
\object{IC\,3568}	&50.0	&4.05	&+	&\MKH\\
\object{NGC\,2392}	&47.0	&3.80	&+	&\MKH\\
\object{NGC\,6629}	&47.0	&3.90	&+	&\MKH\\
\object{EGB\,5}		&42.0	&5.80	&+	&\MKHHG\\
\object{HD\,128220}	&40.6	&4.50	&$-$	&\R\\
\object{IC\,4593}	&40.0	&3.60	&+	&\MKH\\
\object{DdDm\,1}		&37.0	&3.40	&+	&\MMK\\
\object{IC\,418}		&36.0	&3.45	&+	&\MKH\\
\object{He\,2-182}	&36.0	&3.50	&+	&\MKH\\
\object{He\,2-108}	&33.0	&3.30	&+	&\MKH\\
\object{Tc\,1}		&33.0	&3.30	&+	&\MKH\\
\object{M\,1-26}		&33.0	&3.30	&+	&\MKH\\
\object{H\,2-1}		&33.0	&3.35	&+	&\MKH\\
\multicolumn{5}{l|}{O(He) stars}\\
\object{HS\,1522+6615}	&140.0	&5.50	&$-$	&\RDW\\
\object{LoTr\,4}		&120.0	&5.50	&+	&\RDW\\
\object{K\,1-27}		&105.0	&6.50	&+	&\RDW\\
\object{HS\,2209+8229}	&100.0	&6.00	&$-$	&\RDW\\
\object{LSE\,259}	&75.0	&4.40	&$-$	&\HBH\\
\object{LSE\,153}	&70.0	&4.75	&$-$	&\HBH\\
\object{LSE\,263}	&70.0	&4.90	&$-$	&\HBH\\
\object{BD+37$^\circ$442}&60.0	&4.00	&$-$	&\Hus\\
\multicolumn{5}{l|}{PG\,1159 stars}\\
\object{RXJ\,0122.9$-$7521}&180.0&7.50	&$-$	&\WWP\\
\object{H\,1504+65}	&170.0	&8.00	&$-$	&\DWH\\
\object{RXJ\,2117.1+3412}&170.0	&6.00	&+	&\RW\\
\object{NGC\,246}	&150.0	&5.70	&+	&\RW\\
\object{PG\,1144+005}	&150.0	&6.50	&$-$	&\DWH\\
\object{PG\,1520+525}	&150.0	&7.50	&+	&\DrH\\	
\object{PG\,1159$-$035}	&140.0	&7.00	&$-$	&\DrH\\
\object{HS\,2324+3944}	&130.0	&6.20	&$-$	&\DWH\\	
\object{Lo\,4}		&120.0	&5.50	&+	&\DWH\\
\object{PG\,1424+535}	&110.0	&7.00	&$-$	&\DrH\\
\object{HS\,1517+7403}	&110.0	&7.00	&$-$	&\DrH\\
\object{PG\,2131+066}	&95.0	&7.50	&$-$	&\DrH\\
\object{MCT\,0130$-$1937}&90.0	&7.50	&$-$	&\DrH\\
\object{PG\,1707+427}	&85.0	&7.50	&$-$	&\DrH\\
\object{PG\,0122+200}	&80.0	&7.50	&$-$	&\DrH\\
\object{HS\,0704+6153}	&75.0	&7.00	&$-$	&\DrH\\
\multicolumn{5}{l|}{DO white dwarfs}\\
\object{KPD\,0005+5106}	&120.0	&7.00	&$-$	&\DWH\\
\object{PG\,0038+199}	&115.0	&7.50	&$-$	&\DWH\\
\object{PG\,0109+111}	&110.0	&8.00	&?	&\DWH,\WBR\\
\object{PG\,1034+001}	&100.0	&7.50	&$-$	&\DWH\\
\object{HS\,1830+7209}	&100.0	&7.20	&$-$	&\DWH\\
\end{tabular}
\begin{tabular}{l@{ }rl@{ }c@{ }l|}
Name	&$\teff$	&$\log g$ &PN	&Ref.\\ \hline
\object{PG\,0108+101}	&95.0	&7.50	&?	&\DWH\\
\object{MCT\,2148$-$294}	&85.0	&7.50	&$-$	&\DWHR\\
\object{PG\,0046+078}	&73.0	&8.00	&$-$	&\DWH,\Rey\\
\object{PG\,0237+116}	&70.0	&8.00	&$-$	&\DWH\\
\object{RE\,0503$-$289}	&70.0	&7.50	&$-$	&\DWH\\
\object{HS\,0111+0012}	&65.0	&7.80	&$-$	&\DWH\\
\object{Lanning 14}	&58.0	&7.90	&$-$	&\DWH\\
\object{PG\,0929+270}	&55.0	&7.90	&$-$	&\DWHR\\
\object{HZ\,21}		&53.0	&7.80	&$-$	&\DWH\\
\object{PG\,1057$-$059}	&50.0	&7.90	&$-$	&\DWHR\\
\object{HD\,149499\,B}	&49.5	&7.97	&$-$	&\NHJ\\
\object{HS\,0103+2947}	&49.5	&8.00	&$-$	&\DWH\\
\object{PG\,1133+489}	&46.0	&8.00	&$-$	&\DWHR\\
\multicolumn{5}{l|}{DAO white dwarfs}\\
\object{Wray\,17-31}	&69.2	&7.14	&+	&\PRB\\
\object{Ton\,320}	&68.8	&7.68	&?	&\BWB,\TK\\
\object{HS\,2115+1148}	&67.0	&6.90	&$-$	&\DHN\\
\object{Ton\,353}	&66.1	&7.11	&$-$	&\BWB\\
\object{LB\,2}		&65.7	&7.67	&$-$	&\BWB\\
\object{PG\,0834+501}	&60.4	&7.11	&$-$	&\BWB\\
\object{Feige\,55}	&56.7	&7.06	&$-$	&\BWB\\
\object{PG\,0134+181}	&56.4	&7.40	&$-$	&\BWB\\
\object{RE\,1016$-$053}	&56.4	&7.74	&$-$	&\BWB\\
\object{RE\,0720$-$318}	&53.6	&7.64	&$-$	&\BOK\\
\object{PG\,1413+015}	&48.1	&7.69	&$-$	&\BWB\\
\object{PG\,2013+400}	&47.8	&7.69	&$-$	&\BWB\\
\object{PG\,1210+533}	&44.8	&7.89	&$-$	&\BWB\\
\object{PG\,1305$-$017}	&44.4	&7.76	&$-$	&\BWB\\
\multicolumn{5}{l|}{DA white dwarfs}\\
\object{PG\,0948+534}	&126.3	&7.27	&$-$	&\LB\\
\object{HS\,2246+0640}	&98.0	&7.04	&$-$	&\HKH\\
\object{HS\,0615+6535}	&98.0	&7.07	&$-$	&\HKH\\
\object{RE\,1738+665}	&95.3	&7.86	&?	&\FKB,\TK\\
\object{PG\,1342+444}	&78.7	&7.82	&$-$	&\BWB\\
\object{PG\,1622+323}	&77.2	&7.84	&$-$	&\FKB\\
\object{HS\,1749+7145}	&76.9	&7.56	&$-$	&\HKH\\
\object{HS\,1827+7753}	&75.8	&7.68	&$-$	&\HKH\\
\object{RE\,0633+200}	&75.8	&8.40	&$-$	&\FKB\\
\object{PG\,1547+015}	&73.0	&7.63	&$-$	&\FKB\\
\object{RE\,0443$-$034}	&72.3	&8.77	&$-$	&\FKB\\
\object{HS\,2244+0305}	&72.0	&7.78	&$-$	&\HKH\\
\object{Ton\,60}		&69.7	&7.01	&$-$	&\FKB\\
\object{Ton\,21}		&69.7	&7.47	&$-$	&\FKB\\
\object{PG\,1532+033}	&69.5	&7.76	&$-$	&\FKB\\
\object{MCT\,2146$-$4320}&67.9	&7.58	&$-$	&\FKB\\
\object{LB\,1628}	&66.4	&7.75	&$-$	&\FKB\\
\object{RE\,2214-491}	&66.1	&7.38	&$-$	&\FKB\\
\object{PG\,1141+078}	&66.0	&7.57	&$-$	&\BWB\\
\object{KUV\,343$-$7}	&65.8	&7.67	&$-$	&\FKB\\
\object{HS\,0951+3620}	&65.0	&7.66	&$-$	&\HKH\\
\object{RE\,0029$-$632}	&63.7	&7.96	&$-$	&\FKB\\
\object{KPD\,2046+3940}	&63.2	&7.77	&$-$	&\FKB\\
\object{Feige\,24}	&62.7	&7.17	&$-$	&\FKB\\
\object{PG\,2244+031}	&62.3	&7.72	&$-$	&\FKB\\
\object{PG\,2353+026}	&62.0	&7.74	&$-$	&\FKB\\
\object{G\,191\,B2B}	&61.2	&7.49	&$-$	&\FKB\\
\object{HS\,0742+2306}	&60.0	&7.66	&$-$	&\HKH\\
\object{RE\,0558$-$373}	&59.6	&7.44	&$-$	&\FKB\\
\object{RE\,0623$-$374}	&58.2	&7.27	&$-$	&\FKB\\
\object{MCT\,2159$-$4129}&56.4	&7.84	&$-$	&\FKB\\
\object{PG\,1234+481}	&56.4	&7.67	&$-$	&\FKB\\
\object{PG\,1548+405}	&56.2	&7.75	&$-$	&\FKB\\
\object{PG\,2331$-$4731}	&55.8	&8.07	&$-$	&\FKB\\
\object{RE\,2024+200}	&55.8	&7.75	&$-$	&\FKB\\
\object{RE\,0720$-$314}	&55.1	&7.92	&$-$	&\FKB\\
\object{PG\,0836+237}	&54.6	&7.60	&$-$	&\FKB\\
\object{PG\,1123+189}	&54.3	&7.76	&$-$	&\NGS\\
\object{LB\,335}	&53.6	&8.23	&$-$	&\NGS\\
\object{GD\,246}		&53.1	&7.85	&$-$	&\NGS\\
\object{PG\,1657+343}	&53.0	&7.76	&$-$	&\FKB\\
\object{RE\,0620+132}	&52.9	&7.83	&$-$	&\FKB\\
\object{RE\,0957+852}	&51.3	&8.37	&$-$	&\NGS\\
\object{RE\,0457$-$280}	&51.2	&7.72	&$-$	&\NGS\\
\object{HZ\,43}		&50.8	&7.99	&$-$	&\FKB\\
\object{RE\,2116+735}	&50.8	&7.72	&$-$	&\NGS\\
\object{RE\,0550$-$240}	&50.7	&8.07	&$-$	&\FKB\\
\object{PG\,0824+288}	&50.5	&7.43	&$-$	&\FKB\\
\object{PG\,2357+296}	&49.9	&7.60	&$-$	&\FKB\\
\object{RE\,2127$-$221}	&49.8	&7.65	&$-$	&\FKB\\
\object{PG\,1403$-$077}	&49.3	&7.59	&$-$	&\FKB\\
\end{tabular}
\begin{tabular}{l@{ }rl@{ }c@{ }l|}
Name	&$\teff$	&$\log g$ &PN	&Ref.\\ \hline
\object{PG\,1040+451}	&49.2	&7.70	&$-$	&\FKB\\
\object{RE\,1711+664}	&49.0	&8.89	&$-$	&\NGS\\
\object{GD\,8}		&48.7	&7.74	&$-$	&\FKB\\
\object{RE\,0427+740}   &48.6   &7.93   &$-$    &\NGS\\
\object{MCT\,2153$-$4156}&48.2	&7.98	&$-$	&\FKB\\
\object{RE\,1529+483}	&47.7	&7.65	&$-$	&\FKB\\
\object{PG\,1526+013}	&47.3	&7.81	&$-$	&\FKB\\
\object{RE\,2004$-$560}	&47.2	&7.63	&$-$	&\FKB\\
\object{PG\,1642+386}	&47.1	&7.55	&$-$	&\FKB\\
\object{RE\,0003+433}	&46.7	&8.88	&$-$	&\FKB\\
\object{PG\,1232+238}	&46.6	&7.83	&$-$	&\NGS\\
\object{GD\,2}		&46.5	&7.83	&$-$	&\NGS\\
\object{RE\,1746$-$703}	&46.4	&8.97	&$-$	&\FKB\\
\object{GD\,257}		&46.0	&7.67	&$-$	&\FKB\\
\object{PG\,1010+064}	&45.6	&7.77	&$-$	&\FKB\\
\object{PG\,0916+065}	&45.4	&7.73	&$-$	&\BWB\\
\object{KUV\,343-6}	&45.4	&7.39	&$-$	&\FKB\\
\object{RE\,1546$-$364}	&45.2	&8.88	&$-$	&\FKB\\
\object{RE\,2324$-$544}	&45.0	&7.94	&$-$	&\FKB\\
\object{GD\,984}		&44.9	&7.77	&$-$	&\NGS\\
\object{RE\,1800+683}	&44.7	&7.80	&$-$	&\NGS\\
\object{RE\,2156$-$543}	&44.3	&7.91	&$-$	&\FKB\\
\object{RE\,0632$-$050}	&44.1	&8.39	&$-$	&\FKB\\
\object{RE\,1820+580}	&44.1	&7.78	&$-$	&\NGS\\
\object{RE\,0715$-$702}	&43.9	&8.05	&$-$	&\FKB\\
\object{RE\,1032+532}	&43.6	&7.95	&$-$	&\NGS\\
\object{PG\,2303+017}	&43.3	&7.85	&$-$	&\FKB\\
\object{PG\,1648+371}   &42.0	&7.83	&$-$	&\FKB\\
\object{HS\,1950$-$432}	&41.3	&7.85	&$-$	&\FKB\\
\object{RE\,1043+490}	&41.1	&7.94	&$-$	&\NGS\\
\object{PG\,1057+719}	&41.1	&7.84	&$-$	&\NGS\\
\object{RE\,1629+780}	&41.0	&7.92	&$-$	&\NGS\\
\object{PG\,2150+021}	&40.6	&7.76	&$-$	&\FKB\\
\object{PG\,0136+251}	&40.3	&8.93	&$-$	&\FKB\\
\object{Ton\,61}       	&39.8	&7.78	&$-$	&\NGS\\
\object{GD\,394}       	&39.6	&7.94	&$-$	&\FKB\\
\object{GD\,50}		&39.5	&9.07	&$-$	&\NGS\\
\object{GD\,153}		&38.9	&7.78	&$-$	&\NGS\\
\object{RE\,0841+032}	&38.3	&7.75	&$-$	&\NGS\\
\object{RE\,1650+403}	&38.1	&7.97	&$-$	&\NGS\\
\object{PG\,2349+286}	&38.1	&7.78	&$-$	&\FKB\\
\object{RE\,1446+632}	&37.9	&7.79	&$-$	&\NGS\\
\object{PG\,1603+432}	&36.3	&7.85	&$-$	&\FKB\\
\object{RE\,1440+750}	&36.2	&8.87	&$-$	&\NGS\\
\object{RE\,1845+682}	&36.1	&8.23	&$-$	&\NGS\\
\object{PG\,0937+506}	&36.0	&7.69	&$-$	&\NGS\\
\object{RE\,0723$-$274}	&35.9	&7.84	&$-$	&\FKB\\
\object{GD\,659}		&35.8	&7.68	&$-$	&\FKB\\
\object{RE\,1024$-$302}	&35.7	&8.95	&$-$	&\FKB\\
\object{PG\,1026+454}	&35.5	&7.70	&$-$	&\NGS\\
\object{PG\,1636+351}	&35.4	&7.98	&$-$	&\NGS\\
\object{RE\,0605$-$482}	&35.3	&7.84	&$-$	&\FKB\\
\object{KPD\,0416$+$4015}&35.2	&7.75	&$-$	&\FKB\\
\object{Feige\,31}	&35.1	&7.64	&$-$	&\FKB\\
\object{GD\,336}	&34.4	&7.91	&$-$	&\NGS\\
\object{PG\,2120+054}	&34.2	&7.80	&$-$	&\FKB\\
\object{RE\,1943+500}	&34.1	&7.97	&$-$	&\FKB\\
\object{Feige\,93}	&34.0	&7.43	&$-$	&\FKB\\
\object{RE\,0239+500}	&33.8	&8.47	&$-$	&\FKB\\
\object{GD\,80}		&33.5	&8.01	&$-$	&\FKB\\
\object{LB\,1663}	&33.4	&7.85	&$-$	&\FKB\\
\object{RE\,0521$-$102}	&33.2	&8.60	&$-$	&\NGS\\
\object{KPD\,1914+0929}	&33.1	&7.79	&$-$	&\FKB\\
\object{Ton\,210}	&32.9	&7.92	&$-$	&\FKB\\
\object{GD\,71}		&32.7	&7.68	&$-$	&\FKB\\
\object{Ton\,S\,72}	&32.5	&8.09	&$-$	&\FKB\\
\object{PG\,0904+511}	&32.2	&8.11	&$-$	&\NGS\\
\object{PHL\,1400}	&32.1	&8.45	&$-$	&\NGS\\
\object{RE\,0512$-$004}	&31.7	&7.40	&$-$	&\NGS\\
\object{RE\,1019$-$140}	&31.5	&7.92	&$-$	&\NGS\\
\object{RE\,1847$-$221}	&31.5	&8.17	&$-$	&\FKB\\
\object{GD\,38}		&31.3	&7.87	&$-$	&\FKB\\
\object{PG\,1609+631}	&31.0	&8.41	&$-$	&\FKB\\
\object{RE\,0809$-$725}	&30.6	&7.90	&$-$	&\FKB\\
\object{PG\,1125$-$026}	&30.7	&8.24	&$-$	&\NGS\\
\object{PG\,1041+580}	&30.3	&7.81	&$-$	&\NGS\\
\object{GD\,421}		&30.2	&7.63	&$-$	&\FKB\\
\object{RE\,0831$-$534}	&30.2	&8.01	&$-$	&\FKB\\
\object{PG\,1620+647}	&30.2	&7.72	&$-$	&\FKB\\
                &       &       &       &\\
\end{tabular}\\
\end{minipage}
\end{table*}

\end{document}